\documentclass{article}
\usepackage[utf8]{inputenc}
\usepackage{fancyhdr,amsmath,amsthm,amssymb,bm,url,enumerate,float}
\usepackage[super,numbers]{natbib}
\usepackage{bbm}
\usepackage{fullpage}
\usepackage{graphicx,caption}
\usepackage{subcaption}
\usepackage{xcolor}
\usepackage{verbatim}
\usepackage{algorithm}
\usepackage{algpseudocode}
\usepackage{authblk}

\newcommand{\keywords}[1]{\textbf{\textit{Keywords---}} #1}

\newcommand{\E}{\mathbb{E}}

\DeclareMathOperator*{\argmin}{arg\,min}

\title{LEARNER: A Transfer Learning Method for Low-Rank Matrix Estimation}
\author[1]{Sean McGrath}
\author[2]{Cenhao Zhu}
\author[3]{Ryan O'Dea}
\author[4]{Min Guo}
\author[4]{Rui Duan}
\affil[1]{\small Department of Biostatistics, Yale School of Public Health, Connecticut, USA}
\affil[2]{\small Operations Research Center, Massachusetts Institute of Technology, Massachusetts, USA}
\affil[3]{\small Department of Epidemiology, Harvard T.H. Chan School of Public Health, Massachusetts, USA}
\affil[4]{\small Department of Biostatistics, Harvard T.H. Chan School of Public Health, Massachusetts, USA}
\date{}

\begin{document}

\maketitle

\begin{abstract}
Low-rank matrix estimation is a fundamental problem in statistics and machine learning with applications across biomedical sciences, including genetics, medical imaging, drug discovery, and electronic health record data analysis. In the context of heterogeneous data generated from diverse sources, a key challenge lies in leveraging data from a source population to enhance the estimation of a low-rank matrix in a target population of interest. We propose an approach that leverages similarity in the latent row and column spaces between the source and target populations to improve estimation in the target population, which we refer to as LatEnt spAce-based tRaNsfer lEaRning (LEARNER). LEARNER is based on performing a low-rank approximation of the target population data which penalizes differences between the latent row and column spaces between the source and target populations. We present a cross-validation approach that allows the method to adapt to the degree of heterogeneity across populations. We conducted extensive simulations which found that LEARNER often outperforms the benchmark approach that only uses the target population data, especially as the signal-to-noise ratio in the source population increases. We also performed an illustrative application and empirical comparison of LEARNER and benchmark approaches in a re-analysis of summary statistics from a genome-wide association study in the BioBank Japan cohort. LEARNER is implemented in the R package \texttt{learner} and the Python package \texttt{learner-py}.
\end{abstract}


\section{Introduction}
For many biomedical datasets, it is typically believed that the underlying signal structures lie in a lower-dimensional subspace. Consequently, low-rank matrix estimation plays a central role in numerous biomedical applications such as in genetics (e.g., uncovering the genetic basis of complex diseases by integrating genome-wide association studies (GWAS) summary statistics \cite{tanigawa2019components, sakaue2021cross}), medical imaging (e.g., enhancing magnetic resonance imaging (MRI) reconstruction \cite{otazo2015low}), drug discovery (e.g., predicting drug-disease associations \cite{luo2018computational}), and electronic health record (EHR) data analysis (e.g., imputing missing data\cite{xu2023inference}). Common statistical methods for low-rank estimation problems include truncated singular value decomposition (SVD) methods \cite{eckart1936approximation, hansen1987truncated} and their variants and special cases, including principal component analysis (PCA) methods \cite{jolliffe2002principal}, nonnegative matrix factorization methods \cite{lee1999learning}, and other factorization methods \cite{lazzeroni2002plaid}.

With the growing availability of high-dimensional, complex, and heterogeneous datasets, there is an increasing demand for methods that integrate and analyze multiple datasets simultaneously. In biomedical research, for instance, data availability often varies significantly across populations, with certain groups being underrepresented compared to others that are more extensively studied.  
This underrepresentation limits the generalizability and applicability of research findings across diverse populations, reducing the effectiveness of medical interventions and models in underrepresented demographic settings.\cite{west2017genomics, martin2019current, fatumo2022roadmap} To address this challenge, researchers are increasingly exploring ways to leverage data or models from well-studied populations to enhance understanding and improve analysis for populations with limited data. This approach, commonly referred to as transfer learning in machine learning literature \cite{weiss2016survey}, has shown promise in bridging these gaps and enabling more equitable biomedical advancements.\cite{gao2020deep, gu2023commute, li2023targeting} While low-rank estimation problems have been extensively studied, limited attention has been paid to scenarios where data from a target population is limited or of low quality. In such cases, leveraging data from a well-studied source population  might improve the low-rank estimation for the target population, where the challenges lie in how to effectively utilize such auxiliary data while accounting for their inherent differences.

Current research has explored matrix estimation methods in settings with multiple data sources. A number of works have developed multi-source PCA methods in various settings differing based on the assumed similarity between the populations.\cite{oba2007heterogeneous, fan2019distributed, duan2023target, shi2024personalized, li2024knowledge,jalan2025optimal}
Oba et al.\ \cite{oba2007heterogeneous} considered the setting where only the noise level differs between the populations. More recently, some works have considered settings where the populations share the same leading eigenspaces\cite{fan2019distributed, duan2023target} and others have considered settings where the populations share a given number of principal components\cite{li2024knowledge, shi2024personalized}. Several works have studied transfer learning methods for matrix factorization and matrix completion problems. For example, Xu et al.\ \cite{xu2021group} proposed matrix factorization methods in settings where the latent components of the source and target populations are identical except for a small subset of rows. Jalan et al.\ \cite{jalan2025optimal} gave an estimation framework for matrix completion in settings where there is a linear shift in the latent spaces between the source and target populations. A limitation of the aforementioned methods for our setting is that they are based on relatively strong assumptions on the similarity between the populations and may not achieve adaptive information borrowing from the source populations.

We propose an approach called \emph{LatEnt spAce-based tRaNsfer lEaRning} (LEARNER) for improving estimation of a low-rank matrix in target populations that allows for flexible patterns of heterogeneity between the source and target populations. This approach leverages similarity in the latent factors in the underlying low-rank structure between the two populations through a penalized optimization problem, which penalizes differences in the latent row and column spaces between the two populations. We propose a scalable numerical optimization approach. Further, we propose a cross-validation approach to select the appropriate degree of transfer learning between the populations. We also present a tuning-parameter-free approach under certain assumptions on the similarity between the latent spaces of the target and source populations. In Section \ref{sec: methods}, we describe LEARNER and benchmark approaches without transfer learning. We evaluate the performance of LEARNER and benchmark approaches in simulation studies in Section \ref{sec: sim}. We apply LEARNER and compare it to benchmark approaches in a data application based on estimating genetic associations in a Japanese population in Section \ref{sec: application}. We conclude with a discussion in Section \ref{sec: discussion}.

\section{Methods} \label{sec: methods}
\subsection{Model and notation}

Let $\Theta_k \in \mathbb{R}^{p \times q}$ denote the underlying low rank signal matrix in the population $k$, where $k = 0$ corresponds to the target population and $k = 1$ corresponds to the source population. Instead of observing $\Theta_k$ directly,  we assume that we only observe a noisy-version estimate $Y_k$ in population $k$. In our running example, $\Theta_k $ is the underlying association between a set of $p$ genetic variants and $q$ phenotypes, where $Y_k$ is subject to estimation error. 

Consider  the model for $k = 0, 1$
\begin{equation}
    Y_k = \Theta_k + Z_k \label{eq: Y_k} 
\end{equation}
where the $Z_k$ are mean-zero noise matrices with the noise level quantified by $\E\|Z_{k}\|_{op} = \sigma_k$.  We denote the signal strength in each population  by $\theta_k =\lambda_{\text{min}}(\Theta_k)$, where $\lambda_{\text{min}}(\cdot)$ denotes the smallest singular value of a matrix. We focus on a setting where estimating $\Theta_0$ in the target population is more challenging, due to either observing larger noise $\sigma_1\lesssim \sigma_0$ or smaller signal strength $\theta_0\lesssim \theta_1$. In practice, this may be due to factors such as a limited sample size in the original cohort used to derive the summary statistics or greater measurement errors in the target population.

We assume that $\Theta_0$ and $\Theta_1$ have rank $r$ where $r \leq \min\{p, q\}$. This implies that $\Theta_0$ admits the decomposition $\Theta_0 = UV^{\top}$ for some $U \in \mathbb{R}^{p \times r}$ and $V \in \mathbb{R}^{q \times r}$. Under certain identifiability conditions, the columns of $U \in \mathbb{R}^{p \times r}$ and $V \in \mathbb{R}^{q \times r}$ can be interpreted as latent genotypic and phenotypic factors, respectively.\cite{sakaue2021cross} Specifically, $U_{i\ell}$ can be interpreted as a measure of the relative importance of the $i$th genetic variant to the $\ell$th latent genotypic factor. Similarly, $V_{j\ell}$ can be interpreted as a measure of the relative importance of the $j$th phenotype to the $\ell$th latent phenotypic factor. The association between the $i$th variant and the $j$th phenotype in the target population can then be expressed as $\Theta_{0,ij} = \sum_{\ell = 1}^r U_{i\ell}V_{j\ell}$. While the target of inference is $\Theta_0$, measures of the relative importance of the genetic variants and phenotypes to the latent components based on $U$ and $V$ can sometimes be of interest themselves. In the data analysis in Section \ref{sec: application}, we further discuss and illustrate measures of relative importance of the genetic variants and phenotypes to the latent components.

The following subsections describe methods to estimate $\Theta_0$.

\subsection{Approach without transfer learning from the source population}

The \emph{truncated singular value decomposition (SVD)} is a conventional approach to estimate $\Theta_0$ which leverages the low-rank structure of $\Theta_0$ but does not leverage the source population data. The truncated SVD of $Y_0$ with rank $r$, denoted by $W_{0}$, is the best rank $r$ approximation of $Y_0$ in the sense that
\begin{equation*}
    W_{0} = \argmin_{W: \mathrm{rank}(W) = r} \| W - Y_0 \|_F^2.
\end{equation*}
We can express $W_{0}$ by $W_{0} = \hat{U}_{0} \hat{\Lambda}_{0} \hat{V}_{0}^{\top}$ where $\hat{U}_{0} \in \mathbb{R}^{p \times r}$ and $\hat{V}_{0} \in \mathbb{R}^{q \times r}$ are orthonormal matrices and $\hat{\Lambda}_0 \in \mathbb{R}^{r \times r}$ is a diagonal matrix. The columns of $\hat{U}_{0}$ and $\hat{V}_{0}$ are referred to as the left and right singular vectors of $Y_0$, respectively, and the diagonal entries of $\hat{\Lambda}_0$ are referred to as the singular values of $Y_0$.

A crucial question in the truncated SVD approach is how to select the rank $r$. We discuss approaches for rank selection in Section \ref{sec: rank}.

\subsection{Latent space-based transfer learning} \label{sec: lat}

We propose an approach to estimate $\Theta_0$ that leverages (i) the low-rank structure of $\Theta_0$ and (ii) the similarity in the latent row and column subspaces between the source and target populations. This approach uses a rank $r$ approximation of $Y_0$ which penalizes discrepancies in the span of the latent row and column spaces between the source and target populations. By penalizing discrepancies based on the span of the latent row and column spaces rather than based on the singular vectors themselves, this approach can effectively borrow information from the source population under weaker conditions on the similarity between the populations (e.g., not requiring that the order of the singular vectors to be the same between the populations). A detailed description of our approach is below.

First, we estimate the genotypic and phenotypic latent subspaces in the source population, which involves the following notation. Let $U_{k} \Lambda_{k} V_{k}^{\top}$ denote the truncated SVD of $\Theta_k$ with rank $r$. Then $\mathcal{P}(U_{k}) := U_{k}U_{k}^{\top}$ is a projection matrix onto the space of latent genotypic factors in population $k$.  Similarly, $\mathcal{P}(V_k) := V_kV_k^{\top}$ is a projection matrix onto the space of latent phenotypic factors in population $k$. We estimate $\mathcal{P}(U_1)$ and $\mathcal{P}(V_1)$ based on the rank $r$ truncated SVD of $Y_1$ (i.e., by $\mathcal{P}(\hat{U}_1)$ and $\mathcal{P}(\hat{V}_1)$, where $\hat{U}_{1} \hat{\Lambda}_{1} \hat{V}_{1}^{\top}$ denotes the rank $r$ truncated SVD of $Y_1$). 

The LEARNER estimator of $\Theta_0$ is given by $\hat{\Theta}_0^{\mathrm{LEARNER}} = \tilde{U}\tilde{V}^{\top}$ where $(\tilde{U}, \tilde{V})$ is the solution to the following optimization problem
\begin{equation} \label{eq: obj lat}
        \argmin_{U \in \mathbb{R}^{p \times r}, V \in \mathbb{R}^{q \times r}} \big\{ \| U V^{\top} - Y_0 \|_F^2 + \lambda_1\| \mathcal{P}_{\perp}(\hat{U}_{1})U \|_F^2 + \lambda_1\|  \mathcal{P}_{\perp}(\hat{V}_{1})V \|_F^2  + \lambda_2 \| U^{\top} U - V^{\top} V \|_F^2 \big\}
\end{equation}
where $\mathcal{P}_{\perp}(\hat{U}_{1}) = I - \mathcal{P}(\hat{U}_{1})$ and $\mathcal{P}_{\perp}(\hat{V}_{1}) = I - \mathcal{P}(\hat{V}_{1})$. Note that $\mathcal{P}_{\perp}(\hat{U}_{1})$ can be interpreted as the projection matrix onto the orthogonal complement of the space of the latent genotypic factors in the source population; $\mathcal{P}_{\perp}(\hat{V}_{1})$ has an analogous interpretation for the phenotypic factors. Therefore, the second and third terms of (\ref{eq: obj lat}) can be interpreted as penalizing differences in the latent genotypic and phenotypic spaces, respectively, between the target and source populations. The last term in (\ref{eq: obj lat}) balances the sizes of $U$ and $V$ (see Section \ref{sec: lat cv}).

Another way of interpreting (\ref{eq: obj lat}) is that it interpolates between the SVDs of $Y_0$ and $Y_1$. In particular, the truncated SVD of $Y_0$ is a special case of LEARNER when $\lambda_1 = 0$ and the truncated SVD of $Y_1$ is a special case when $\lambda_1 = \infty$. In this regard, LEARNER adapts to the level of heterogeneity between the two populations when using suitable $\lambda_1, \lambda_2$ values. 

Note that we let the second and third terms have identical penalties (i.e., $\lambda_1$). In cases where one believes that the degree of similarity between the latent genotypic factors between the two populations is highly different than the degree of similarity between the latent phenotypic factors between the populations, one can use different penalties on the second and third terms. Further details are given in the supplementary materials.

In the following subsections, we outline how to numerically solve (\ref{eq: obj lat}), how to adapt the approach in the presence of missing data, how to select $\lambda_1$ and $\lambda_2$, and how to select the rank.

\subsubsection{Numerical optimization approach} \label{sec: lat cv}

The objective function (\ref{eq: obj lat}) is non-convex which is common in the low-rank matrix estimation literature. To numerically solve (\ref{eq: obj lat}), we adopt an alternating minimization strategy which has been commonly employed to solve similar nonconvex matrix optimization problems \cite{hardt2014understanding, zhang2018predicting, li2021scmfmda}. Given an initial point, this approach updates $U$ by minimizing the objective function (via a gradient descent step) treating $V$ as fixed. Then, in the same manner, $V$ is updated treating $U$ as fixed. These updates of $U$ and $V$ are repeated until convergence. 

Algorithm \ref{alg: lat} summarizes the numerical optimization approach used to solve (\ref{eq: obj lat}). Throughout Algorithm \ref{alg: lat}, $f$ denotes the objective function in (\ref{eq: obj lat}). For the stopping criteria, we terminate the for loop whenever any of the following conditions occur: (i) the value of objective function does not significantly change between iterations (i.e., $|\epsilon_t -\epsilon_{t-1}|$ is sufficiently small), (ii) a maximum number of iterations is reached, or (iii) the value of the objective function begins to diverge (e.g., $\epsilon_t > 10\epsilon_0$). The implementation of Algorithm \ref{alg: lat} involves computing gradients of $f$ with respect to $U$ and $V$. A straightforward calculation shows that these gradients are given by
\begin{align*}
    \nabla_{U} f(U,V) & = 2(UV^{\top}V - Y_0V) + 2 \lambda_1 \mathcal{P}_{\perp}(\hat{U}_{1})U + 4 \lambda_2 U(U^{\top}U - V^{\top}V) \\
    \nabla_{V} f(U,V) & = 2(VU^{\top}U - Y_0^{\top}U) + 2 \lambda_1 \mathcal{P}_{\perp}(\hat{V}_{1})V + 4 \lambda_2 V(V^{\top}V - U^{\top}U).
\end{align*}

Note that regularization term $\lambda_2 \| U^{\top} U - V^{\top} V \|_F^2$ is included in the objective function to help avoid local minima caused by scale ambiguity in $U$ and $V$. That is, $UV^\top$ is unchanged by the re-scaling $(U,V) \rightarrow (\frac{1}{c}U, cV)$ for a nonzero constant $c$. Consequently, $\lambda_2$ can be interpreted as a tuning parameter that helps balance the norms of $U$ and $V$ to mitigate such local minima.

Although the objective function is non-convex, recent work in the low-rank matrix estimation literature has provided support for local minimizers to such problems. Under suitable regularity conditions, recent theoretical work has found that local minima are very close to or exactly equal the global minimum in spite of non-convexity.\cite{zhao2015nonconvex, ge2016matrix, ge2017no, li2019non} Empirically, local search methods have often been found to be effective and computationally efficient and have consequently seen widespread use.\cite{koren2009bellkor, koren2009matrix} We investigate the performance of our approach in simulations and empirical evaluations.

\begin{algorithm}  [h!]
\textbf{Input} Target population data $Y_0 \in \mathbb{R}^{p \times q}$, source population data $Y_1 \in \mathbb{R}^{p \times q}$, rank $r \in \mathbb{N}$, regularization parameters $\lambda_1, \lambda_2 \in \mathbb{R^+}$, step size $c \in \mathbb{R^+}$\\
\textbf{Output} Estimate $\hat{\Theta}_0^{\mathrm{LEARNER}} \in \mathbb{R}^{p \times q}$\\
\textbf{function} $\textrm{LEARNER}(Y_0, Y_1, r, \lambda_1, \lambda_2, c)$
\caption{LEARNER}
\label{alg: lat}
\begin{algorithmic}[1] 
\State Initialize $U^{(0)} \in \mathbb{R}^{p \times r}, V^{(0)} \in \mathbb{R}^{q \times r}$ based on the SVD of $Y_1$
\State Initialize $\epsilon_{0} \leftarrow f(U^{(0)}, V^{(0)})$
\For{$t = 1, 2, \dots$}
    \State Update $U^{(t)} \leftarrow U^{(t-1)} - c \frac{\| U^{(t-1)} \|_F}{\|\nabla_{U} f(U^{(t-1)},V^{(t-1)})\|_F}  \nabla_{U} f(U^{(t-1)},V^{(t-1)})$
    \State Update $V^{(t)} \leftarrow V^{(t-1)} - c \frac{\| V^{(t-1)} \|_F}{\|\nabla_{V} f(U^{(t)},V^{(t-1)})\|_F}  \nabla_{V} f(U^{(t)},V^{(t-1)})$
    \State Update $\epsilon_{t} \leftarrow f(U^{(t)}, V^{(t)})$
\EndFor
\State Set $t_{\mathrm{best}} = \argmin_t \epsilon_t$
\State Set $\hat{\Theta}_0^{\mathrm{LEARNER}} = U^{(t_{\mathrm{best}})} {V^{(t_{\mathrm{best}})}}^{\top}$
\end{algorithmic}
\end{algorithm}

\subsubsection{Handling missing data} \label{sec: lat missing data}

When $Y_0$ has missing data, we can adapt LEARNER as follows. Let $\Omega$ denote the set of the indices of the non-missing entries in $Y_0$. We can solve the following optimization problem
\begin{align} 
        \argmin_{U, V} \Bigg\{ & \frac{1}{|\Omega| / pq}\sum_{(i,j) \in \Omega}( (U V^{\top})_{ij} - Y_{0,ij} )^2 + \lambda_1\| \mathcal{P}_{\perp}(\hat{U}_{1})U \|_F^2 + \lambda_1\|  \mathcal{P}_{\perp}(\hat{V}_{1})V \|_F^2 + \lambda_2 \| U^{\top} U - V^{\top} V \|_F^2 \bigg\} \label{eq: obj lat missing}
\end{align}
However, it will be convenient to re-express (\ref{eq: obj lat missing}) as follows
\begin{align} \label{eq: obj lat missing v2}
    \argmin_{U, V} \big\{ & \frac{1}{|\Omega| / pq} \| UV^{\top} - \tilde{Y}_0 \|_F^2 + \lambda_1\| \mathcal{P}_{\perp}(\hat{U}_{1})U \|_F^2 + \lambda_1\|  \mathcal{P}_{\perp}(\hat{V}_{1})V \|_F^2 + \lambda_2 \| U^{\top} U - V^{\top} V \|_F^2 \big\}
\end{align}
where $\tilde{Y}_{0}$ is given by
\begin{equation*}
    (\tilde{Y}_{0})_{ij} = \begin{cases}
        (Y_0)_{ij} \quad & \text{if $(i,j) \in \Omega$} \\ (U V^{\top})_{ij} & \text{otherwise}.
    \end{cases}
\end{equation*}
We can then apply Algorithm \ref{alg: lat} to solve (\ref{eq: obj lat missing v2}), where the first term in $\nabla_{U} f(U,V)$ and $\nabla_{V} f(U,V)$ is now scaled by $\frac{1}{|\Omega| / pq}$.

Note that scaling the first term in these optimization problems by the percentage of non-missing entries of $Y_0$ becomes important when applying cross-validation to select $\lambda_1$ and $\lambda_2$ (see Section \ref{sec: transfer learning}). This scaling allows the relative size of the first term to remain the same when varying the percentage of missing entries in $Y_0$, such as when holding out one fourth of the entries in $Y_0$ in cross-validation versus when not holding out any entries in $Y_0$ in the application with the selected $\lambda_1$ and $\lambda_2$.

\subsubsection{Selecting the degree of transfer learning} \label{sec: transfer learning}

The penalties $\lambda_1, \lambda_2$ control how much information is borrowed from the source population when estimating $\Theta_0$. We consider a cross-validation approach to select $\lambda_1, \lambda_2$ so that the method can adapt to the level of heterogeneity between the source and target populations and thus can protect against negative transfer. More specifically, cross-validation enables LEARNER to increasingly leverage the source population data to estimate $\Theta_0$ as (i) the degree of overlap in the spaces of the latent factors between the source and target populations increases and (ii) the signal strength in the source population increases.

We consider the following four-fold cross-validation approach to select $\lambda_1$ and $\lambda_2$. To form the training and test datasets, we randomly partition the entries $Y_0$ into four equally sized subsamples. The training datasets are obtained by removing one of the four subsamples and the corresponding test datasets are based on the held out subsamples. In fold $k$, let $\Omega_k$ denote the indices of $Y_0$ used for the training set and $\Omega_k^{\perp}$ denote the indices of $Y_0$ used for the test set.

We consider a grid of values for $\lambda_1$ and $\lambda_2$, denoted by $\mathcal{S}_1$ and $\mathcal{S}_2$. For each $(\lambda_1, \lambda_2) \in \mathcal{S}_1 \times \mathcal{S}_2$, we apply LEARNER to the training datasets and evaluate their mean squared errors (MSEs) on the test datasets. That is, in validation set $k$, the MSE is given by
\begin{equation}
    \text{MSE}(\lambda_1, \lambda_2, k) = \frac{1}{|\Omega_k^{\perp}|} \sum_{(i,j) \in \Omega_k^{\perp}} (\hat{\Theta}^{\mathrm{LEARNER},\lambda_1,\lambda_2,k}_{ij} - Y_{0,ij})^2. \label{eq: mse cv}
\end{equation}
where $\hat{\Theta}^{\mathrm{LEARNER},\lambda_1,\lambda_2,k}$ denote the LEARNER estimate from training set $k$. We select the value of $(\lambda_1, \lambda_2) \in \mathcal{S}_1 \times \mathcal{S}_2$ with the smallest MSE across all four validation datasets (i.e. $\frac{1}{4}\sum_{k = 1}^4\text{MSE}(\lambda_1, \lambda_2, k)$).

\subsubsection{Selecting the rank} \label{sec: rank}

Rank selection in matrix estimation problems has been studied by a number of seminal works in the theoretical and methodological literature (see Donoho et al.\ \cite{donoho2023screenot} and references within). For LEARNER, we consider estimating $r$ by applying the ScreeNOT \cite{donoho2023screenot} method to $Y_1$ since we assume that the target and the source populations share the same rank. ScreeNOT selects an optimal value for hard thresholding the singular values of $Y_1$. That is, for the hard thresholding estimator $W_{1,\theta}:=\sum_{i:\hat{\lambda}_i >\theta} \hat{\lambda}_i \hat{u}_i \hat{v}_i^{\top}$, this approach selects the value of $\theta$ that minimizes $\| W_{1,\theta} - \Theta_1\|_F^2$. Note that this approach allows for various noise structures, such as correlated noise across the rows and/or columns of $Y_1$. In situations where the source population and the target population have distinct underlying ranks, $r_0$ and $r_1$, respectively, rank estimation can be performed separately for each population. However, the difference in ranks introduces additional dissimilarity between the latent spaces, making the source data less useful, particularly when $r_1 < r_0$.

\subsubsection{Summary}

Algorithm \ref{alg: lat full} summarizes the complete LEARNER method, including the selection of the rank $r$ and regularization parameters $\lambda_1$ and $\lambda_2$.

\begin{algorithm}  [h!]
\textbf{Input} Target population data $Y_0 \in \mathbb{R}^{p \times q}$, source population data $Y_1 \in \mathbb{R}^{p \times q}$, set of candidate $\lambda_1$ values $\mathcal{S}_1$, set of candidate $\lambda_2$ values $\mathcal{S}_2$, step size $c \in \mathbb{R}^+$\\
\textbf{Output} Estimate $\hat{\Theta}_0^{\mathrm{LEARNER}} \in \mathbb{R}^{p \times q}$
\caption{LEARNER with Nuisance Parameter Selection}
\label{alg: lat full}
\begin{algorithmic}[1] 
\State Apply ScreeNOT to $Y_1$ to select $r$
\For{$k = 1, 2, 3, 4$}
        \State Select training dataset indices $\Omega_k$
        \State Set $Y_0^{\mathrm{train},k}$ by setting entries corresponding to $\Omega_k$ to missing values
    \EndFor
\For{$(\lambda_1, \lambda_2) \in \mathcal{S}_1 \times \mathcal{S}_2$}
    \For{$k = 1, 2, 3, 4$}
        \State Set $\hat{\Theta}_{0}^{\mathrm{LEARNER},\lambda_1,\lambda_2,k} = \text{LEARNER}(Y_0^{\mathrm{train},k}, Y_1, r, \lambda_1, \lambda_2, c)$
        \State Set $\text{MSE}(\lambda_1, \lambda_2, k)$ by applying equation (5) in the main text
    \EndFor
    \State Set $\text{MSE}(\lambda_1, \lambda_2) = \frac{1}{4}\sum_{k = 1}^4\text{MSE}(\lambda_1, \lambda_2, k)$
\EndFor
\State Set $(\lambda_1^{(\mathrm{best})}, \lambda_2^{(\mathrm{best})}) = \argmin_{(\lambda_1, \lambda_2)} \text{MSE}(\lambda_1, \lambda_2)$
\State Set $\hat{\Theta}_{0}^{\mathrm{LEARNER}} = \text{LEARNER}(Y_0, Y_1, r, \lambda_1^{(\mathrm{best})},  \lambda_2^{(\mathrm{best})}, c)$
\end{algorithmic}
\end{algorithm}

\subsection{Direct projection LEARNER} \label{sec: dlat}

When we have prior knowledge about the latent spaces of the genotypic and phenotypic factors are the same between the source and target populations (i.e., $\mathcal{P}(U_0) = \mathcal{P}(U_1)$ and $\mathcal{P}(V_0) = \mathcal{P}(V_1)$), we can consider a simpler estimation method than LEARNER. This approach estimates $\Theta_0$ by directly projecting $Y_0$ onto the genotypic and phenotypic latent spaces learned from the source population, which we refer to as the \emph{direct projection LEARNER (D-LEARNER)} approach. We can express this estimator as $$\hat{\Theta}_0^{\mathrm{D-LEARNER}} = \mathcal{P}(\hat{U}_1) Y_0 \mathcal{P}(\hat{V}_1).$$

A key advantage of this approach is that it does not require selecting the tuning parameters $\lambda_1$ and $\lambda_2$, which can be computationally expensive and statistically challenging in some settings (e.g., see Section \ref{sec: sim correlated}).  However, the projection method forces the target estimate to conform to the singular spaces of the source, potentially increasing errors when the source population diverges significantly from the target. This alignment may not accurately reflect the characteristics of the target population, leading to distorted or misleading results.

\section{Simulation study} \label{sec: sim}

\subsection{Independent noise} \label{sec: sim independent noise}

Our first set of simulations considers the setting of independent noise. These simulations evaluate how the performance of the LEARNER, D-LEARNER, and the truncated SVD approach are affected by the (i) similarity between the latent spaces of the source and target populations, (ii) signal-to-noise ratio in the source population, (iii) matrix rank $r$, and (iv) the matrix dimensions $(p,q)$ in the setting of independent noise. To characterize similarity between the latent spaces of the source and target populations in these simulations, we define the measures
\begin{align*}
   d_U & := \| \mathcal{P}(U_1) - \mathcal{P}(U_0)\|_F \\
   d_V & := \| \mathcal{P}(V_1) - \mathcal{P}(V_0)\|_F.
\end{align*}

\subsubsection{Simulation design} \label{sec: sim design correct spec}

We set $\Theta_0, \Theta_1 \in \mathbb{R}^{p \times q}$, where we considered a rectangular setting with $(p, q) = (5000, 50)$ and a square setting with $(p,q) = (500,500)$. We considered $r \in \{4, 8\}$. We set $\Theta_0$ by generating a $p \times q$  matrix with i.i.d.\ $\text{Normal}(0,1)$ entries and then performing a truncated SVD with rank $r$. We considered three scenarios for $\Theta_1$ based on the similarity of the latent spaces of the source and target populations:
\begin{enumerate}
    \item \textit{High similarity}: We let $\Theta_1$ have the same left and right singular vectors as $\Theta_0$ but in different order. Specifically, we set $\Theta_1$ by reversing the order of the singular values of $\Theta_0$. Note that $d_U = d_V = 0$ in this case. 
    \item \textit{Moderate similarity}: We set $\Theta_1$ by reversing the order of the singular values of $\Theta_0$ and adding perturbations to the left and right singular vectors of $\Theta_0$. Specifically, we set the left singular vector matrix of $\Theta_1$ by adding a matrix with i.i.d. $\text{Uniform}(-\frac{1}{8\sqrt{p}}, \frac{1}{8\sqrt{p}})$ entries to the left singular vector matrix of $\Theta_0$ and orthonormalizing the resulting matrix. Similarly, we added i.i.d. $\text{Uniform}(-\frac{1}{8\sqrt{q}}, \frac{1}{8\sqrt{q}})$ perturbations to the right singular vector matrix of $\Theta_0$ and then orthonormalized the resulting matrix. This resulted in $d_U \approx 0.40, d_V \approx 0.42$ when $r = 4$ and $d_U \approx 0.57, d_V \approx 0.54$ when $r = 8$ in the rectangular matrix setting; In the square matrix setting, $d_U \approx 0.40, d_V \approx 0.41$ when $r = 4$ and $d_U \approx 0.57, d_V \approx 0.57$ when $r = 8$.
    \item \textit{Low similarity}: We set $\Theta_1$ in the same manner as in the moderate similarity setting, except that the perturbations to the left singular vectors of $\Theta_0$ were generated by $\text{Uniform}(-\frac{1}{2\sqrt{p}}, \frac{1}{2\sqrt{p}})$ and the perturbations to the right singular vectors were generated by $\text{Uniform}(-\frac{1}{2\sqrt{q}}, \frac{1}{2\sqrt{q}})$. This resulted in $d_U \approx 0.78, d_V \approx 0.83$ when $r = 4$ and $d_U \approx 1.11, d_V \approx 1.05$ when $r = 8$; In the square matrix setting, $d_U \approx 0.78, d_V \approx 0.80$ when $r = 4$ and $d_U \approx 1.10, d_V \approx 1.10$ when $r = 8$.
\end{enumerate}
We simulated $Y_0$ and $Y_1$ by (\ref{eq: Y_k}) with i.i.d. normal noise. To evaluate the effect of increasing the signal-to-noise ratio in the source population, we set $\sigma_0^2 = 0.1$ and considered $\sigma_1^2 \in \{\frac{\sigma_0^2}{10}, \frac{\sigma_0^2}{5}, \frac{\sigma_0^2}{3}, \sigma_0^2\}$. 

In summary, there were $3 \times 4 \times 2 \times 2 = 48$ different data generating scenarios by varying the similarity in the latent spaces of the source and target populations (3 levels), the variance of the noise in the source population (4 levels, $\sigma_1^2 \in \{\frac{\sigma_0^2}{10}, \frac{\sigma_0^2}{5}, \frac{\sigma_0^2}{3}, \sigma_0^2\}$), and rank (2 levels, $r\in \{4, 8\}$), and matrix dimensions (2 levels, $(p,q) \in \{(5000,50), (500,500)\}$). In each scenario, we performed 50 repetitions.

We compared the performance of the truncated SVD, LEARNER, and D-LEARNER methods. D-LEARNER and LEARNER estimated the rank using ScreeNOT. To serve as a benchmark, the true rank was used in the truncated SVD method. Details on the hyperparameter values used for LEARNER (e.g., candidate $\lambda_1$ and  $\lambda_2$ values) in these simulations are in the supplementary material. For each method, we evaluated the average Frobenius norm of the estimation error (i.e., $\| \hat{\Theta}_0 - \Theta_0 \|_F$) across the 50 repetitions.

\subsubsection{Results}

\begin{figure} [h!]
    \centering
    \includegraphics[width=0.75\textwidth]{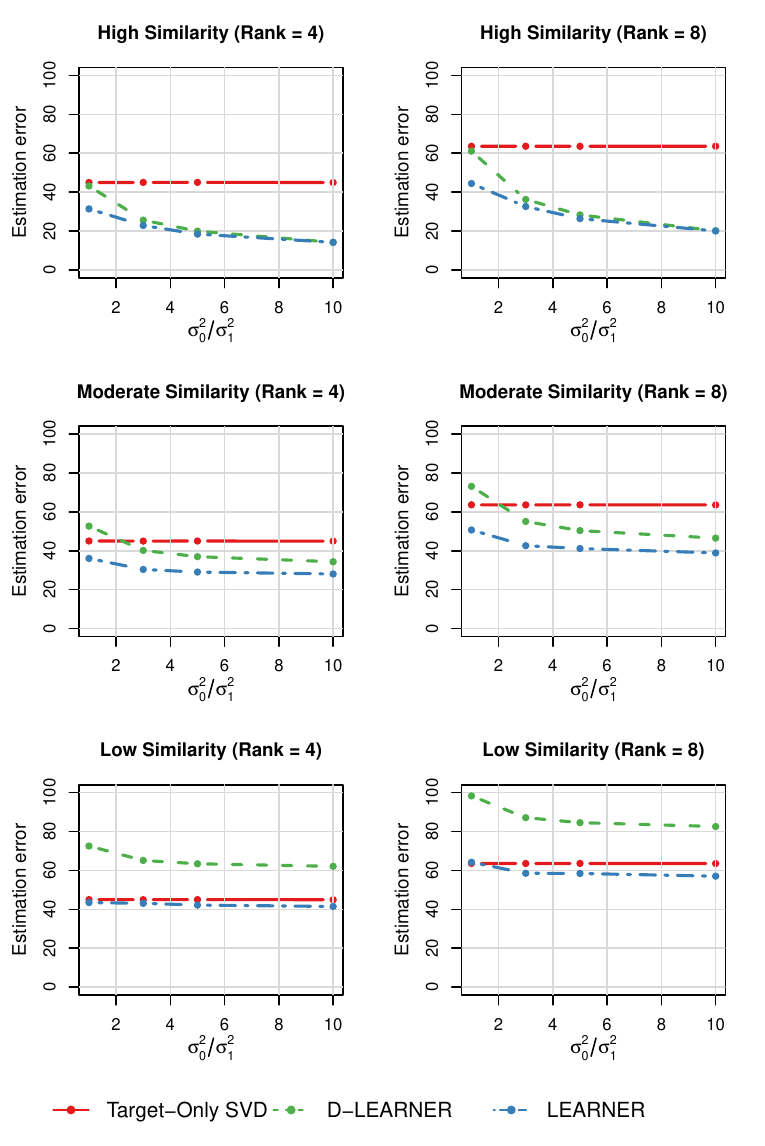}
    \caption{Simulation results in the rectangular matrix settings with independent noise.\label{fig: independent noise}}
\end{figure}

The simulation results in the rectangular matrix settings are illustrated in Figure \ref{fig: independent noise}. In the high similarity simulation scenarios, the LEARNER and D-LEARNER methods performed best and similarly to each other. The performance of these two methods improved as the noise variance in the source population decreased. 

Similar trends held in the moderate similarity scenario. LEARNER and D-LEARNER effectively leveraged the source population data to improve estimation of $\Theta_0$; however, as one may expect, the degree of improvement for these methods -- especially for D-LEARNER -- was less compared to the scenarios with high similarity in the latent spaces. 

The low similarity scenarios represent cases where only very limited information can be leveraged from the source population. As such, LEARNER only had a small improvement over the target-only truncated SVD approach. Since D-LEARNER assumes that the latent spaces between the source and target populations are identical, this approach performed worse than the target-only truncated SVD approach in these scenarios. 

Trends were very similar between the rank 4 scenarios and the corresponding rank 8 scenarios. The estimation error of all three approaches that assume a low rank structure increased as the rank increased. The rank selection method used by LEARNER and D-LEARNER correctly selected the rank in each iteration in these simulations. 

The same trends held in the square matrix setting with $(p,q) = (500,500)$. These results are given in  the supplementary materials. Note that, unlike the simulation settings with correlated noise (see Section \ref{sec: sim correlated}), we do not include error bars in the figures illustrating the standard deviation of the estimation error in the independent noise settings because they are too small to be visible.

\subsection{Correlated noise} \label{sec: sim correlated}

Our next set of simulations explores how the performance of these methods are affected by correlated noise.

\subsubsection{Simulation design}

We set $\Theta_0$ the same as in the simulations with independent noise (i.e., Section \ref{sec: sim design correct spec}), and we set $\Theta_1$ as in the high similarity, moderate similarity, and low similarity scenarios described in the independent noise simulations. We fixed the rank $r = 4$ and the matrix dimensions $(p,q)=(5000,50)$.

We simulated $Y_0$ and $Y_1$ by equations (\ref{eq: Y_k}), although we now consider correlated noise settings. Letting $Z_{0,j}^{\top}$ and $Z_{1,j}^{\top}$ denote the $j$th columns of $Z_0$ and $Z_1$ respectively, we considered that $Z_{0,j}^{\top} \stackrel{\mathrm{i.i.d.}}{\sim} \text{Normal}(0, \Sigma_0)$ and $Z_{1,j}^{\top} \stackrel{\mathrm{i.i.d.}}{\sim} \text{Normal}(0, \Sigma_1)$. We let $\Sigma_0$ and $\Sigma_1$ have exchangeable covariance structures with a correlation of $\rho$ ($\rho \in \{0.1, 0.25, 0.5\}$). That is, we let $\Sigma_{0,j_1,j_2} = \rho^{I(j_1 \neq j_2)}\sigma_0^2 $ and $\Sigma_{1,j_1,j_2} = \rho^{I(j_1 \neq j_2)}\sigma_1^2 $. As in the simulations with independent noise, we fixed $\sigma_0^2 = 0.1$ and considered $\sigma_1^2 \in \{\frac{\sigma_0^2}{10}, \frac{\sigma_0^2}{5}, \frac{\sigma_0^2}{3}, \sigma_0^2\}$. 

In summary, there were $3 \times 3 \times 4 = 36$  different data generating scenarios by varying the similarity in the latent spaces of the source and target populations (3 levels), correlation ($\rho \in \{0.1, 0.25, 0.5\}$) and variance of the noise in the source population ($\sigma_1^2 \in \{\frac{\sigma_0^2}{10}, \frac{\sigma_0^2}{5}, \frac{\sigma_0^2}{3}, \sigma_0^2\}$).

We applied the same methods as described in the simulations with independent noise, and evaluated the average Frobenius norm of the estimation error across the 50 repetitions. The supplementary material contains details on the hyperparameter values used by LEARNER in these simulations.

\subsubsection{Results}

\begin{figure} [h!]
    \centering
    \includegraphics[width=0.95\textwidth]{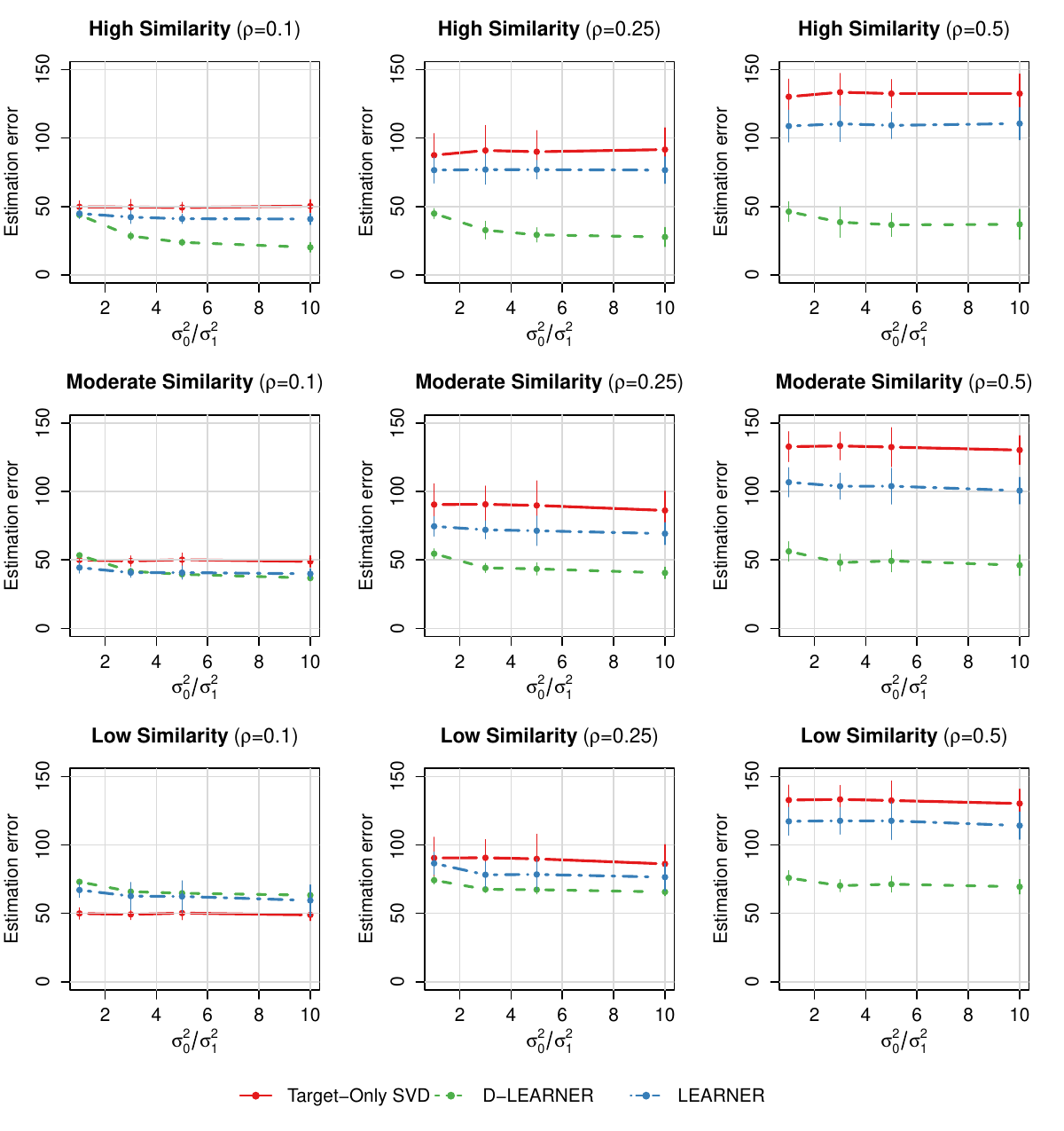}
    \caption{Simulation results under the correlated noise scenarios. The error bars correspond to the standard deviation of the estimation error across the 50 repetitions. \label{fig: correlated noise same}}
\end{figure}

The simulation results in the settings with high similarity in the latent spaces are summarized in the top row of Figure \ref{fig: correlated noise same}. D-LEARNER performed best in these scenarios, especially when the correlation was large. Its estimation error was not strongly affected by the degree of correlation. While LEARNER also improved on the target-only SVD approach, it did not perform as well as D-LEARNER. This may be attributed to selecting too small $\lambda_1$ values due to the fact that the held-out data in cross-validation was correlated with the non-held-out data. The rank selection method used by LEARNER and D-LEARNER incorrectly selected a rank of 5 (recall that the true rank was 4) in each iteration in each of the scenarios. 

Similar trends held in the simulation scenarios with moderate and low similarity. LEARNER and D-LEARNER often outperformed the target-only SVD approach. However, the correlation in the noise had a bigger impact on LEARNER than D-LEARNER, which is likely due to the cross-validation approach in LEARNER performing sub-optimally.

In the supplementary material, we describe a variation of LEARNER that selects the tuning parameters $(\lambda_1, \lambda_2)$ based on using an independent dataset from the target population rather than by using held-out entries of $Y_0$. We find that this variation of LEARNER performs considerably better in correlated noise settings, generally outperforming the target-only SVD and D-LEARNER methods.

\section{Data application} \label{sec: application}

To illustrate an application of LEARNER and D-LEARNER, we obtained publicly available GWAS summary statistics from BioBank Japan (BBJ), UK Biobank (UKB), and FinnGen.\cite{sakaue2021cross} BBJ is a biobank containing clinical and genomic data of participants who were recruited from 12 medical institutes in Japan based on having a diagnosis of one of 47 diseases.\cite{nagai2017overview} UKB is a biobank with over 500,000 individuals from 22 assessment centers in the United Kingdom.\cite{bycroft2018uk}. FinnGen combines data from Finland’s national health register (comprising all Finnish residents since 1969) with a number of Finnish biobanks.\cite{kurki2023finngen} In our analyses, we consider the BBJ population to be the target population and combine the UKB European population and FinnGen European population to be the source population. 

We selected disease endpoints (which we refer to as phenotypes throughout) and genetic variants in a manner generally consistent with that of Sakaue et al.\ \cite{sakaue2021cross}. We identified 145 phenotypes available in both the source and target populations. We then performed a standard screening procedure to select genetic variants significant in at least one of the datasets and at least one of the phenotypes. Additionally, genetic variants that are in high linkage disequilibrium (LD) were removed. We identified 25,415 genetic variants and 145 disease endpoints. The matrix $Y_1$ was formed by performing an inverse-variance weighted meta-analysis of the genetic associations across the UKB and FinnGen populations. For 27 of the phenotypes, source population data were only available from the UKB data and not the FinnGen data, in which case no meta-analysis was performed for these phenotypes. In total, the sample size for estimating the genetic associations was around 179,000 in the target population and 628,000 in the source population. A total of 0.006\% of the entries in $Y_0$ were missing, and 0.105\% of the entries in $Y_1$ were missing.
See the supplementary material for further details on the data processing. 

The original analysis in Sakaue et al.\ \cite{sakaue2021cross} concatenated the observed matrices in the source and target populations and applied the truncated SVD to the resulting matrix to estimate the latent representation of genetic variants and phenotypes. By combining the populations in this manner, this analysis treats different phenotypes within a population equivalently as phenotypes across different populations. Here, we adopt a transfer learning approach by applying the LEARNER and D-LEARNER methods to estimate $\Theta_0$.

\subsection{Exploratory data analyses}

In exploratory data analyses, we compared the latent components between the source and target populations based on $Y_0$ and $Y_1$. These analyses focus on the projection matrices onto the latent phenotypic and genotypic spaces (i.e., $\mathcal{P}(\hat{U}_k)$ and $\mathcal{P}(\hat{V}_k)$ where $\hat{U}_k$ and $\hat{V}_k$ are the orthonormal left and right singular vector matrices, respectively, of $Y_k$) as well as the so-called phenotype and variant contribution scores \cite{sakaue2021cross}. The phenotype contribution score quantifies the relative importance of a given phenotype for a given latent component. Specifically, the phenotype contribution score of the $i$th phenotype for the $\ell$th latent component of $Y_k$ is given by $\hat{V}_{k, i, \ell}^2$, which ranges from 0 to 1. Similarly, the variant contribution score quantifies the relative importance of a given variant for a given latent component, which is defined as $\hat{U}_{k, i, \ell}^2$ for the $i$th variant and $\ell$th latent component of $Y_k$.

To select the number of latent components in our analyses, we applied ScreeNOT to $Y_1$ resulting in a rank of 6. We then computed the truncated SVD of $Y_0$ and $Y_1$ with the selected rank to compute the projection matrices and the phenotype and variant contribution scores. 

\begin{figure}[h!]
    \centering
    \begin{subfigure}{0.4\textwidth}
       \centering
       \includegraphics[width=\textwidth]{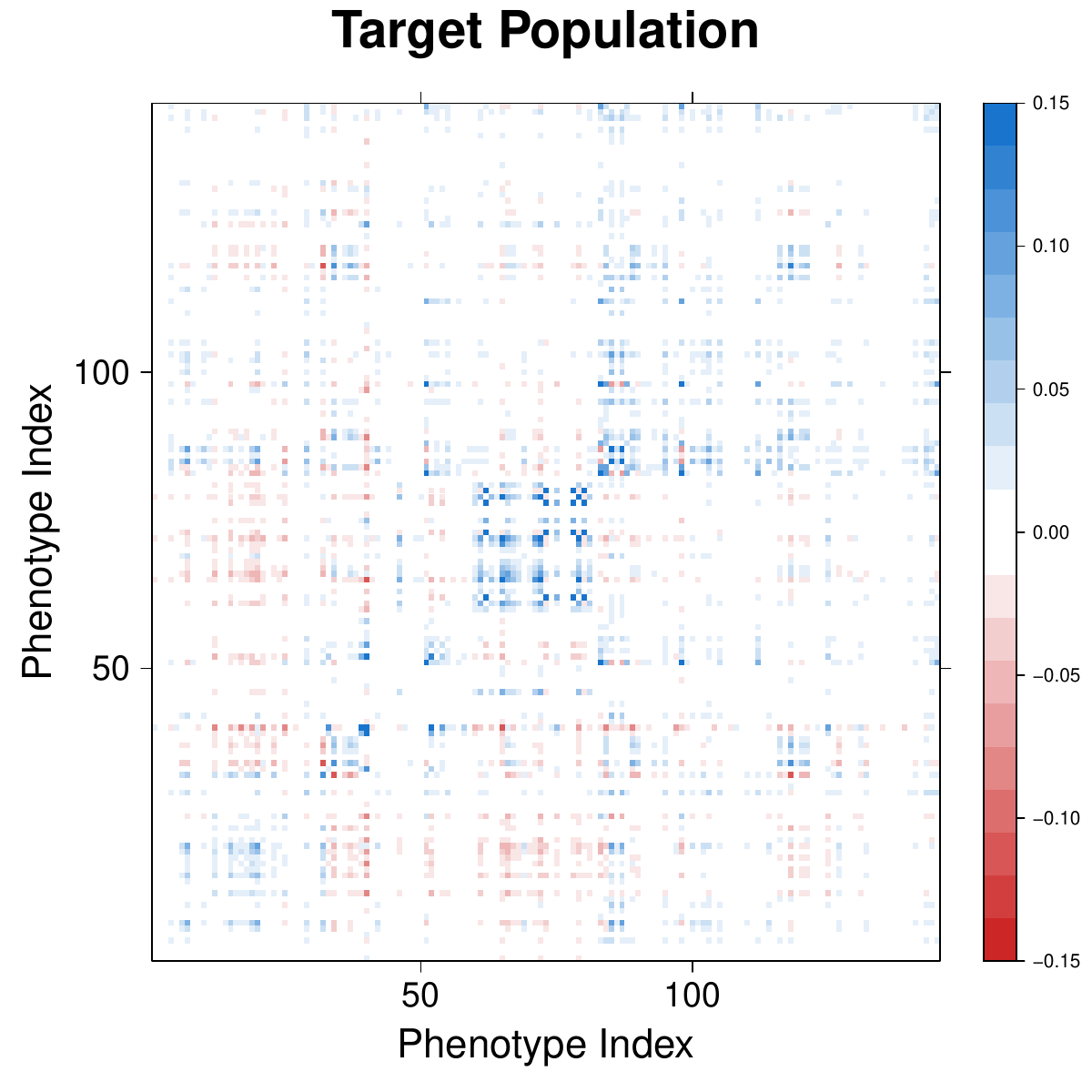}
  \end{subfigure}%
    ~ 
    \begin{subfigure}{0.4\textwidth}
        \centering
        \includegraphics[width=\textwidth]{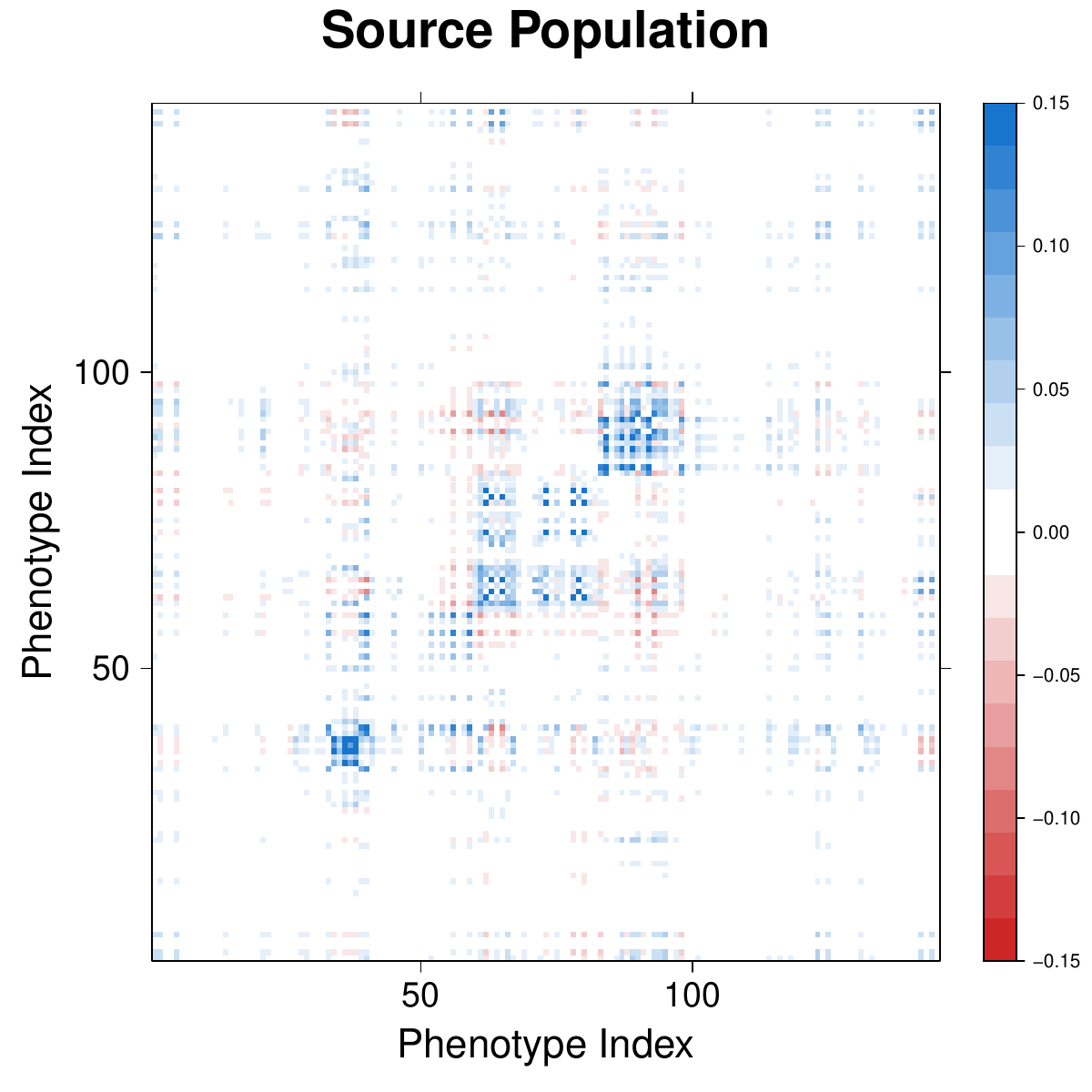}
    \end{subfigure}
    \begin{subfigure}{0.4\textwidth}
        \centering
        \includegraphics[width=\textwidth]{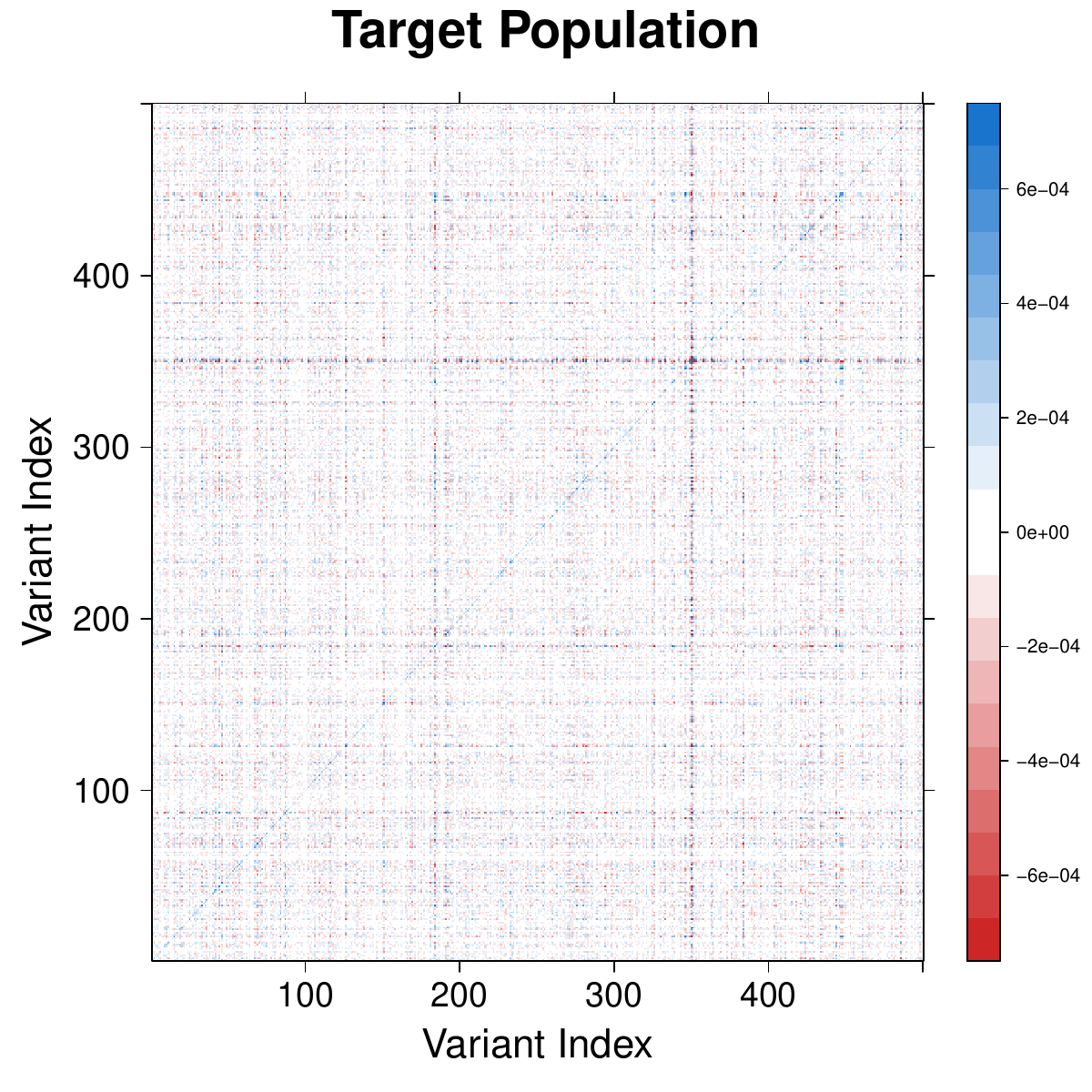}
    \end{subfigure}%
    ~ 
    \begin{subfigure}{0.4\textwidth}
        \centering
        \includegraphics[width=\textwidth]{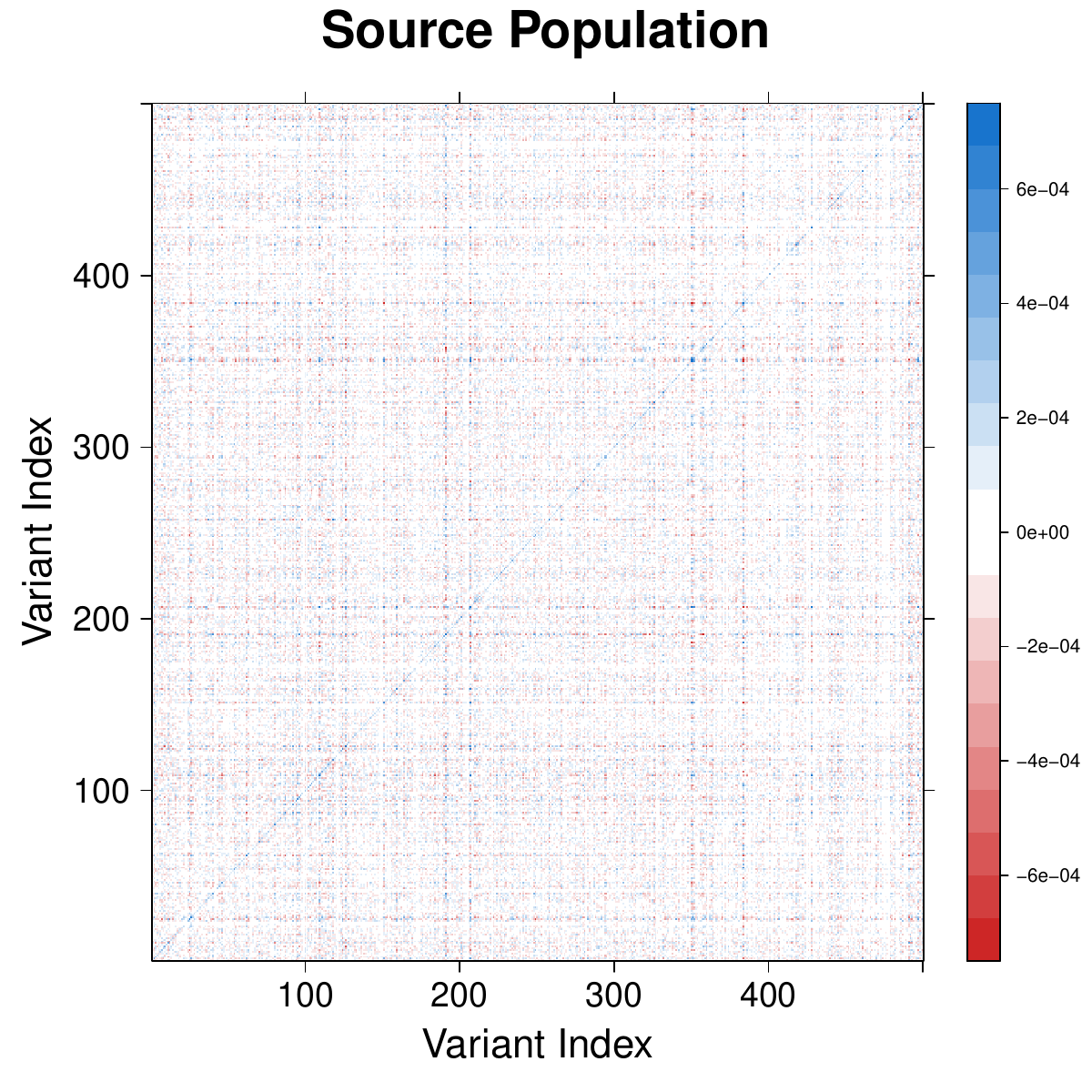}
    \end{subfigure}
    \caption{Heatmaps of $\mathcal{P}(\hat{V}_0)$ (top left panel) and $\mathcal{P}(\hat{V}_1)$ (top right panel) and 500 randomly selected subset of rows and columns of $\mathcal{P}(\hat{U}_0)$ (bottom left panel) and $\mathcal{P}(\hat{U}_1)$ (bottom right panel). The phenotypes are ordered based on their ICD-10 category, and the variants are ordered based on their chromosome and position number.\label{fig: PU PV}}
\end{figure}

The projection matrices for the latent phenotypic components are given in the top row of Figure \ref{fig: PU PV}. There is moderate similarity between the target and source populations. The phenotype contribution scores in the two populations are illustrated in the supplementary material. We also describe the top phenotypes (defined as phenotypes with the highest contribution scores) in the latent components in the target and source populations in the supplementary material. 
Most latent phenotypic components in the source and target populations could be characterized by well-known disease categories such as cardiovascular disease, pulmonary disorders, thyroid disorders. There were a number of similar latent phenotypic components in both the source and target populations. For example, the first latent component in each population was characterized by angina pectoris (with top phenotypes of angina pectoris, stable angina pectoris, unstable angina pectoris). There were also clear differences in the latent phenotypic factors between the two populations. The source population had a latent component characterized by diabetes (type 1 diabetes, type 2 diabetes), whereas the target population had a latent component characterized by thyroid disorders (hypothyroidism, hyperthyroidism, Hashimoto's disease). Each population also had a latent component whose top phenotypes did not fall into a well-defined disease category. Specifically, the fourth latent component in the target population had top phenotypes of hypothyroidism, angina pectoris, allergic rhinitis, and pneumonia, and the fifth latent component in the source population had top phenotypes of cerebral aneurysm, rheumatoid arthritis, ischemic stroke, and Graves' disease. We elaborate on the interpretation of the latent components when analyzing the LEARNER and D-LEARNER estimates in the following subsection.  

The bottom row of Figure \ref{fig: PU PV} compares the genotypic latent spaces between the target and source populations. The variant contribution scores are illustrated in the supplementary material. In both populations, the variant contribution scores were relatively evenly distributed across the chromosome and position numbers. Both populations shared several top variants (defined as variants with the highest contribution scores). For example, variant 9:22119195 (chr:pos) had the highest contribution score in the first latent component in both the source and target populations.

\subsection{LEARNER illustration}

We applied LEARNER to estimate $\Theta_0$ as described in Section \ref{sec: lat}. The supplementary material describes the hyperparameter used, the results of the analyses selecting the tuning parameters, and a heatmap of the LEARNER estimate of $\Theta_0$.

\begin{figure}[h!]
    \centering
    \begin{subfigure}{0.4\textwidth}
        \centering
        \includegraphics[width=\textwidth]{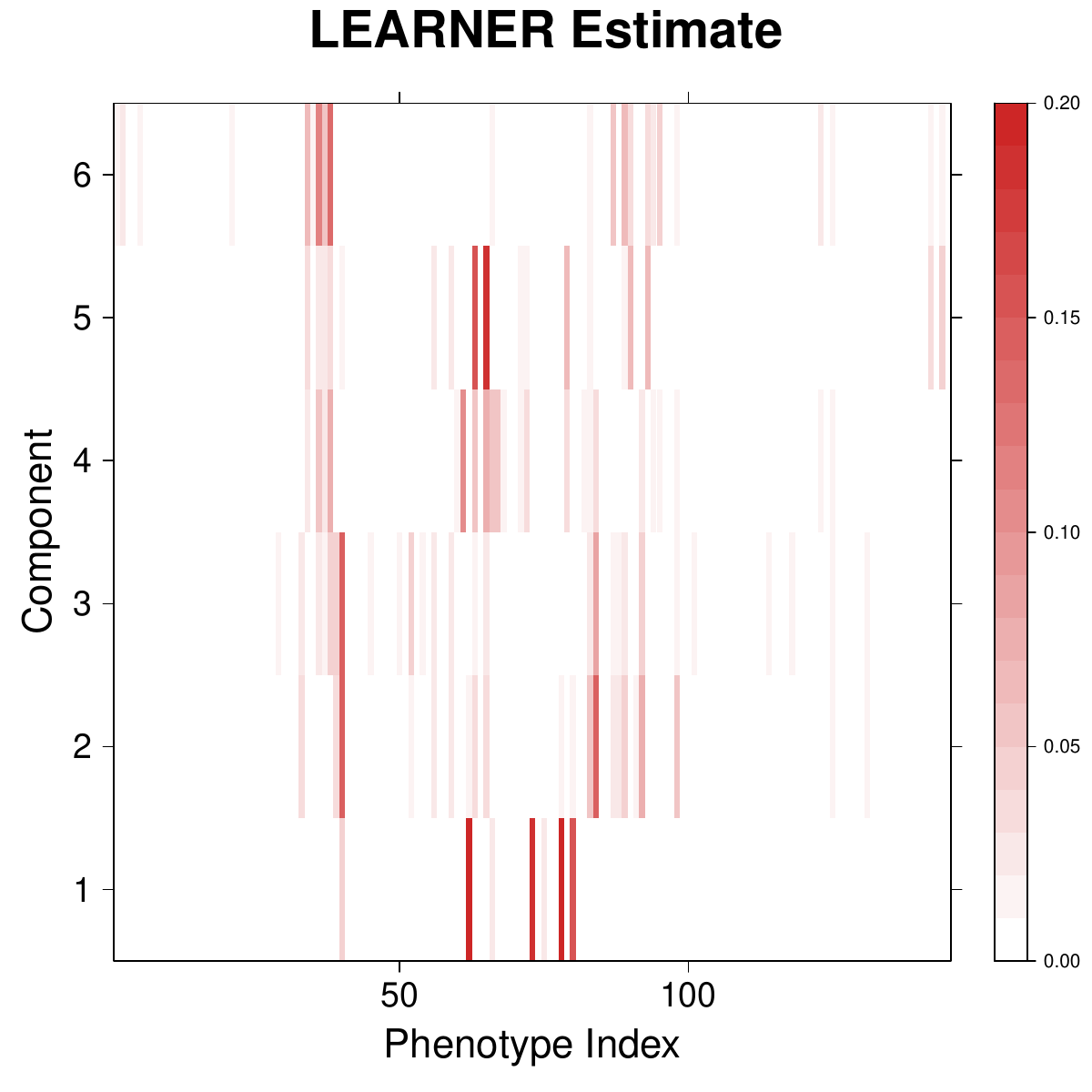}
    \end{subfigure}%
    ~ 
    \begin{subfigure}{0.4\textwidth}
        \centering
        \includegraphics[width=\textwidth]{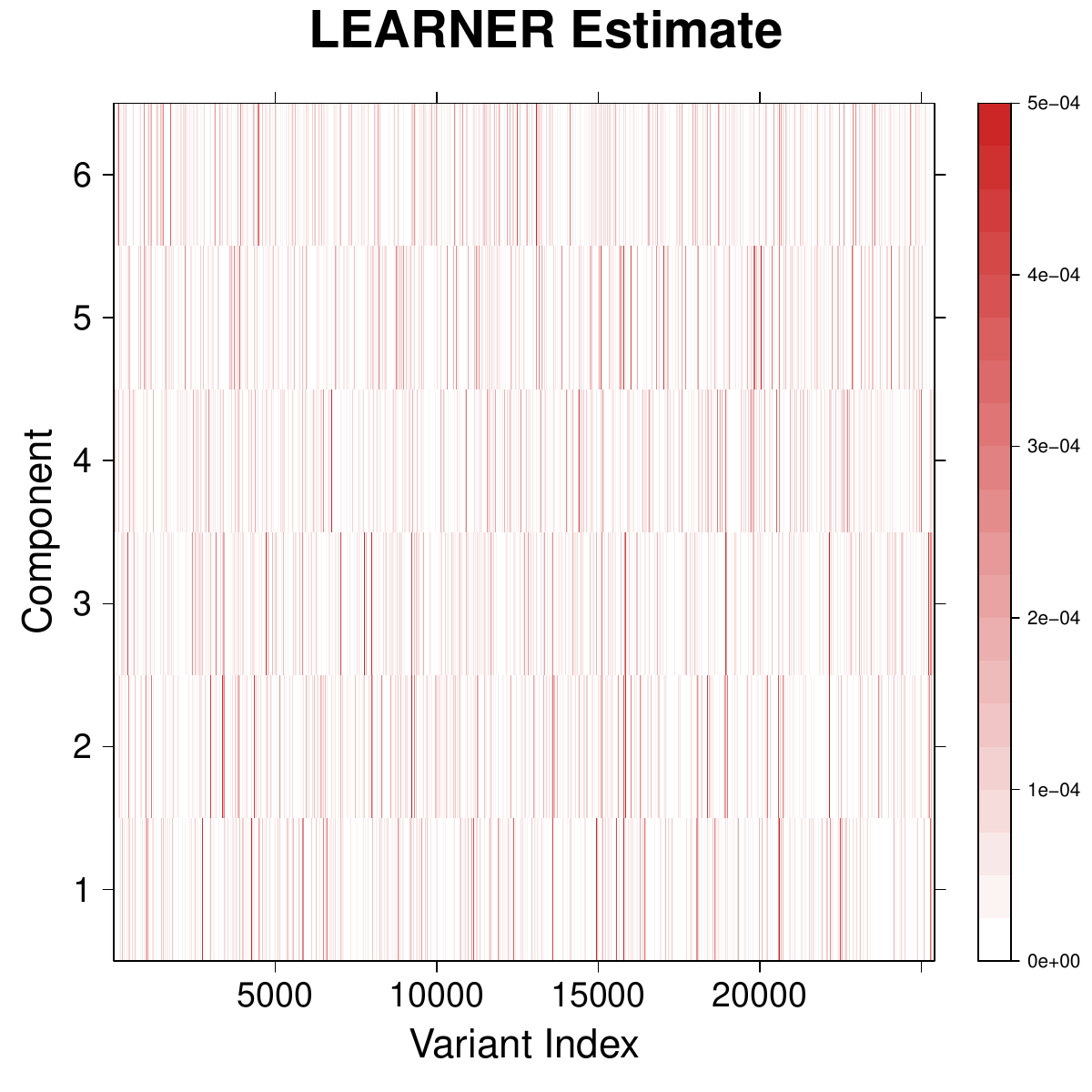}
    \end{subfigure}  
    \begin{subfigure}{0.4\textwidth}
    \end{subfigure}
    
    \caption{Heatmaps of the matrix of phenotype (left panel) and variant (right panel) contribution scores in the target population based on LEARNER. The phenotypes are ordered based on their ICD-10 category, and the variants are ordered based on their chromosome and position number.\label{fig: contribution learner}}
\end{figure}

The top row of Figure \ref{fig: contribution learner} illustrates the phenotype and variant contribution scores in the target population based on LEARNER. To compute the contribution scores, we applied the singular value decomposition to the LEARNER estimate of $\Theta_0$. The top phenotypes in each latent component are listed in Table \ref{tab: key phenotypes target learner}. Each latent factor could be characterized by one or two disease classes. While there were similarities in the phenotype contribution scores obtained from LEARNER and those in the target-only truncated SVD approach (e.g., components characterized by angina and diabetes), there were also some key differences. LEARNER had latent components characterized by thyroid disorders, aneurysm, and pulmonary disorders, which were found in the source population but not the target population. The variant contribution scores were relatively evenly distributed, as found in the target-only and source-only truncated SVD analysis. As one may expect, top variants in the source and target-only analyses such as 9:22119195 were top variants for LEARNER as well.

\begin{table}[h!]
\caption{Top phenotypes in each latent factor in the target population based on LEARNER. For each latent factor, we list the four phenotypes with the highest contribution scores. The phenotype contribution scores are in parentheses. The phenotypes related to the latent factor characterization are in bold text.\label{tab: key phenotypes target learner}}
\begin{center}
\begin{tabular}{p{0.06\textwidth}p{0.25\textwidth}p{0.60\textwidth}} \hline 
            Factor & Characterization & Top phenotypes   \\ \hline
            1 & Angina & \textbf{Angina pectoris (0.26)}, \textbf{stable angina pectoris (0.23)}, myocardial infarction (0.18), \textbf{unstable angina pectoris (0.16)} \\
            2 & Diabetes/Asthma & \textbf{Type 2 diabetes (0.15)}, \textbf{asthma (0.14)}, \textbf{pediatric asthma (0.07)}, allergic rhinitis (0.06) \\
            3 & Diabetes/Asthma & \textbf{Type 2 diabetes (0.14)}, \textbf{asthma (0.08)}, hypothyroidism (0.04), \textbf{pediatric asthma (0.04)}  \\
            4 & Cardiovascular Disease & \textbf{Atrial flutter/fibrillation (0.10)}, \textbf{cerebral aneurysm (0.07)}, hypothyroidism (0.07), \textbf{chronic heart failure (0.06)} \\ 
            5 & Aneurysm & \textbf{Cerebral aneurysm (0.19)}, \textbf{unruptured cerebral aneurysm (0.15)}, interstitial lung disease (0.07), pulmonary Fibrosis (0.06) \\
            6 & Thyroid Disorders & \textbf{Hypothyroidism (0.13)}, \textbf{Hashimoto's disease (0.12)}, chronic obstructive pulmonary disease (0.06), \textbf{Graves' disease (0.06)} \\ \hline
\end{tabular}
\end{center}
\end{table}

To further interpret the latent phenotypic factors, we present the top phenotypes after performing a varimax rotation \cite{kaiser1958varimax} to right singular matrix estimated by LEARNER in the supplementary materials. Unlike the original (unrotated) analyses, the latent components with the varimax rotation showed more distinct characterizations, with no two components being defined by the same disease class.

We also applied D-LEARNER to estimate $\Theta_0$ and performed analogous analyses for this approach. There were many similarities between the results from the LEARNER and D-LEARNER analyses, although the latent components from D-LEARNER were more similar to the source-only analysis. These results are detailed in the supplementary materials.

\subsection{Empirical evaluation}

We performed 5-fold cross-validation analyses to compare the performance of the LEARNER, D-LEARNER, and truncated SVD approaches in the context of the data application.  The training and test sets were obtained by randomly partitioning the entries of $Y_0$. We applied LEARNER to the training data as described in Section \ref{sec: lat}. We also applied a missing value SVD approach -- also referred to as hard thresholding \cite{mazumder2010spectral} -- to the (target population) training data. This approach numerically solves the following optimization problem
\begin{equation*}
        W_{0,r} := \argmin_{W: \mathrm{rank}(W) = r}  \sum_{(i,j) \in \Omega_k}( W_{ij} - Y_{0,ij} )^2 
\end{equation*}
where $\Omega_k$ denotes the set of the indices corresponding to the $k$th training dataset. We used the \texttt{softImpute} R package \cite{softImpute} to apply this approach. We applied the D-LEARNER approach as described in Section \ref{sec: dlat}, using $W_{0,r}$ in place of $Y_0$. Last, we applied the truncated SVD to the source population data ($Y_1$) to estimate $\Theta_0$. We used a rank of 6 for all approaches. For each approach, we computed the MSE as well as the 2.5th and 97.5th percentiles of the squared errors in the corresponding validation data.

The complete results are given in the supplementary material. LEARNER performed the best in each of the test sets, followed by D-LEARNER, the target-only truncated SVD, and the source-only truncated SVD. The relatively small differences in the estimation errors between all the approaches may be attributed to using noisy ``true" values when computing the MSE (i.e., using the held out entries of $Y_{0}$ rather than $\Theta_{0}$).

Although LEARNER demonstrates empirical superiority over D-LEARNER in the holdout dataset, further external validation would be valuable, such as a larger meta-analysis of these genetic associations in an East Asian population or an evaluation from a disease risk prediction perspective using an external dataset from the same population.

\begin{table}[h]
\caption{Results of the cross-validation analyses comparing LEARNER, D-LEARNER, and the truncated SVD approaches. Results include the mean squared error (MSE) and the 2.5th and 97.5th percentiles of the squared errors ($Q_{(2.5,97.5)}$). \label{tab: cv}}
\begin{center}
\begin{tabular}{lllllllll} \hline
             & \multicolumn{2}{c}{LEARNER} & \multicolumn{2}{c}{D-LEARNER} & \multicolumn{2}{c}{Target-Only SVD} & \multicolumn{2}{c}{Source-Only SVD}   \\ \cline{2-3}  \cline{4-5} \cline{6-7} \cline{8-9}
            Test Set & MSE & $Q_{(2.5,97.5)}$ & MSE & $Q_{(2.5,97.5)}$ & MSE & $Q_{(2.5,97.5)}$ & MSE & $Q_{(2.5,97.5)}$   \\ \hline
            1 & 1.079 & (0.001, 5.550)  & 1.092 & (0.001, 5.621) & 1.106 & (0.001, 5.742) & 1.220 & (0.001, 6.434)   \\
            2 & 1.071 &  (0.001, 5.524) & 1.084 & (0.001, 5.585) & 1.101 & (0.001, 5.721) & 1.213 & (0.001, 6.392) \\
            3 & 1.071 &  (0.001, 5.526) & 1.085 & (0.001, 5.599) & 1.101 & (0.001, 5.718) & 1.212 & (0.001, 6.392) \\
            4 & 1.072 &  (0.001, 5.540) & 1.085 & (0.001, 5.609) & 1.102 & (0.001, 5.742) & 1.213 & (0.001, 6.415) \\
            5 & 1.070 & (0.001, 5.526) & 1.083 & (0.001, 5.587) & 1.101 & (0.001, 5.739) & 1.212 & (0.001, 6.384)  \\ \hline
\end{tabular}
\end{center}
\end{table}

\section{Discussion} \label{sec: discussion}

We proposed an approach called LEARNER that leverages data from a source population to improve estimation of a low-rank matrix in a target population. This approach penalizes differences between the latent row and column spaces between the source and target populations. Further, we presented a cross-validation approach that enables LEARNER to select the appropriate degree of information borrowing from the source population and consequently protects against negative transfer. Our simulation and empirical evaluations illustrated that LEARNER can effectively improve inference in the target population under various settings. In particular, our simulation results showed that LEARNER generally performed better than the benchmark approach that only uses the target population data, especially as the similarity in the latent spaces increased and as the signal strength in the source population increased. We also presented a tuning-parameter-free approach called D-LEARNER for the special case where the latent row and column spaces are believed to be the same between the target and source populations. These approaches are implemented in the R package \verb|learner|, available on CRAN at \url{https://CRAN.R-project.org/package=learner}, and the Python package \verb|learner-py|, available on PyPI at \url{https://pypi.org/project/learner-py/}.

Our data application focused on the problem of estimating genetic associations across different phenotypes, which can ultimately be used in disease risk prediction as well as in clustering diseases and genetic variants.\cite{sakaue2021cross, tanigawa2019components} Although underrepresentation of target populations and heterogeneity across populations are well-known problems in this literature, \cite{fatumo2022roadmap, martin2019current} statistical methods adopting a transfer learning approach have not been applied in this context to the best of our knowledge. 

Several modifications to the proposed methodology could enhance the flexibility or performance of LEARNER's application.
First, as we discussed in the supplementary materials, column- and row-space specific penalties can be added when the degrees of similarity in the latent row and column spaces between the source and target populations are believed to be highly different. Second, the rank in each population can be selected separately when needed.
Additionally, one may consider extensions for the setting with multiple different source populations. For example, under the assumption that the source populations have the same latent spaces, one could pool information across the sources to better estimate the projection matrices onto the latent factors. Indeed, Shi and Kontar \cite{shi2024personalized} and Li et al.\ \cite{li2024knowledge} adopted similar ideas in multi-source PCA contexts. Fourth, adaptations to account for heteroskedastic noise may be considered when estimating latent subspaces or determining rank. Previous research indicates that heteroskedastic noise can substantially affect the accuracy of conventional spectral decomposition methods.\cite{zhang2022heteroskedastic} Incorporating techniques that address heteroskedasticity into the LEARNER algorithm could improve its performance.\cite{zhang2022heteroskedastic,landa2025dyson}

\subsection*{Acknowledgments}
The simulations and data application in this work were run on the FASRC Cannon cluster supported by the FAS Division of Science Research Computing Group at Harvard University.

\subsection*{Financial disclosure}

Sean McGrath was supported by the National Science Foundation Graduate Research Fellowship Program under Grant No.\ DGE2140743. Rui Duan was supported by National Institutes of Health (R01 GM148494).

\subsection*{Conflict of interest}

The authors declare no potential conflict of interests.

\subsection*{Open Research}
The code for the simulation study and data application, along with links to download the source data, is publicly available at  \url{https://github.com/stmcg/learner-paper}.

\bibliographystyle{unsrt}
\bibliography{ref-cur}

\begin{thebibliography}{10}

\bibitem{tanigawa2019components}
Yosuke Tanigawa, Jiehan Li, Johanne~M Justesen, Heiko Horn, Matthew Aguirre,
  Christopher DeBoever, Chris Chang, Balasubramanian Narasimhan, Kasper Lage,
  Trevor Hastie, et~al.
\newblock Components of genetic associations across 2,138 phenotypes in the uk
  biobank highlight adipocyte biology.
\newblock {\em Nature communications}, 10(1):4064, 2019.

\bibitem{sakaue2021cross}
Saori Sakaue, Masahiro Kanai, Yosuke Tanigawa, Juha Karjalainen, Mitja Kurki,
  Seizo Koshiba, Akira Narita, Takahiro Konuma, Kenichi Yamamoto, Masato
  Akiyama, et~al.
\newblock A cross-population atlas of genetic associations for 220 human
  phenotypes.
\newblock {\em Nature genetics}, 53(10):1415--1424, 2021.

\bibitem{otazo2015low}
Ricardo Otazo, Emmanuel Candes, and Daniel~K Sodickson.
\newblock Low-rank plus sparse matrix decomposition for accelerated dynamic mri
  with separation of background and dynamic components.
\newblock {\em Magnetic resonance in medicine}, 73(3):1125--1136, 2015.

\bibitem{luo2018computational}
Huimin Luo, Min Li, Shaokai Wang, Quan Liu, Yaohang Li, and Jianxin Wang.
\newblock Computational drug repositioning using low-rank matrix approximation
  and randomized algorithms.
\newblock {\em Bioinformatics}, 34(11):1904--1912, 2018.

\bibitem{xu2023inference}
Zhiwei Xu, Ziming Gan, Doudou Zhou, Shuting Shen, Junwei Lu, and Tianxi Cai.
\newblock Inference of dependency knowledge graph for electronic health
  records.
\newblock {\em arXiv preprint arXiv:2312.15611}, 2023.

\bibitem{eckart1936approximation}
Carl Eckart and Gale Young.
\newblock The approximation of one matrix by another of lower rank.
\newblock {\em Psychometrika}, 1(3):211--218, 1936.

\bibitem{hansen1987truncated}
Per~Christian Hansen.
\newblock The truncated svd as a method for regularization.
\newblock {\em BIT Numerical Mathematics}, 27:534--553, 1987.

\bibitem{jolliffe2002principal}
Ian~T Jolliffe.
\newblock {\em Principal component analysis for special types of data}.
\newblock Springer, 2002.

\bibitem{lee1999learning}
Daniel~D Lee and H~Sebastian Seung.
\newblock Learning the parts of objects by non-negative matrix factorization.
\newblock {\em nature}, 401(6755):788--791, 1999.

\bibitem{lazzeroni2002plaid}
Laura Lazzeroni and Art Owen.
\newblock Plaid models for gene expression data.
\newblock {\em Statistica sinica}, pages 61--86, 2002.

\bibitem{west2017genomics}
Kathleen~McGlone West, Erika Blacksher, and Wylie Burke.
\newblock Genomics, health disparities, and missed opportunities for the
  nation’s research agenda.
\newblock {\em Jama}, 317(18):1831--1832, 2017.

\bibitem{martin2019current}
Alicia~R Martin, Masahiro Kanai, Yoichiro Kamatani, Yukinori Okada, Benjamin~M
  Neale, and Mark~J Daly.
\newblock Current clinical use of polygenic scores will risk exacerbating
  health disparities.
\newblock {\em Nature genetics}, 51(4):584, 2019.

\bibitem{fatumo2022roadmap}
Segun Fatumo, Tinashe Chikowore, Ananyo Choudhury, Muhammad Ayub, Alicia~R
  Martin, and Karoline Kuchenbaecker.
\newblock A roadmap to increase diversity in genomic studies.
\newblock {\em Nature medicine}, 28(2):243--250, 2022.

\bibitem{weiss2016survey}
Karl Weiss, Taghi~M Khoshgoftaar, and DingDing Wang.
\newblock A survey of transfer learning.
\newblock {\em Journal of Big data}, 3:1--40, 2016.

\bibitem{gao2020deep}
Yan Gao and Yan Cui.
\newblock Deep transfer learning for reducing health care disparities arising
  from biomedical data inequality.
\newblock {\em Nature communications}, 11(1):5131, 2020.

\bibitem{gu2023commute}
Tian Gu, Phil~H Lee, and Rui Duan.
\newblock Commute: communication-efficient transfer learning for multi-site
  risk prediction.
\newblock {\em Journal of biomedical informatics}, 137:104243, 2023.

\bibitem{li2023targeting}
Sai Li, Tianxi Cai, and Rui Duan.
\newblock Targeting underrepresented populations in precision medicine: A
  federated transfer learning approach.
\newblock {\em The Annals of Applied Statistics}, 17(4):2970--2992, 2023.

\bibitem{oba2007heterogeneous}
Shigeyuki Oba, Motoaki Kawanabe, Klaus-Robert M{\"u}ller, and Shin Ishii.
\newblock Heterogeneous component analysis.
\newblock {\em Advances in Neural Information Processing Systems}, 20, 2007.

\bibitem{fan2019distributed}
Jianqing Fan, Dong Wang, Kaizheng Wang, and Ziwei Zhu.
\newblock Distributed estimation of principal eigenspaces.
\newblock {\em Annals of statistics}, 47(6):3009, 2019.

\bibitem{duan2023target}
Junting Duan, Markus Pelger, and Ruoxuan Xiong.
\newblock Target pca: Transfer learning large dimensional panel data.
\newblock {\em Journal of Econometrics}, page 105521, 2023.

\bibitem{shi2024personalized}
Naichen Shi and Raed~Al Kontar.
\newblock Personalized pca: Decoupling shared and unique features.
\newblock {\em Journal of Machine Learning Research}, 25(41):1--82, 2024.

\bibitem{li2024knowledge}
Zeyu Li, Kangxiang Qin, Yong He, Wang Zhou, and Xinsheng Zhang.
\newblock Knowledge transfer across multiple principal component analysis
  studies.
\newblock {\em arXiv preprint arXiv:2403.07431}, 2024.

\bibitem{jalan2025optimal}
Akhil Jalan, Yassir Jedra, Arya Mazumdar, Soumendu~Sundar Mukherjee, and
  Purnamrita Sarkar.
\newblock Optimal transfer learning for missing not-at-random matrix
  completion.
\newblock {\em arXiv preprint arXiv:2503.00174}, 2025.

\bibitem{xu2021group}
Kan Xu, Xuanyi Zhao, Hamsa Bastani, and Osbert Bastani.
\newblock Group-sparse matrix factorization for transfer learning of word
  embeddings.
\newblock In {\em International Conference on Machine Learning}, pages
  11603--11612. PMLR, 2021.

\bibitem{hardt2014understanding}
Moritz Hardt.
\newblock Understanding alternating minimization for matrix completion.
\newblock In {\em 2014 IEEE 55th Annual Symposium on Foundations of Computer
  Science}, pages 651--660. IEEE, 2014.

\bibitem{zhang2018predicting}
Wen Zhang, Xiang Yue, Weiran Lin, Wenjian Wu, Ruoqi Liu, Feng Huang, and Feng
  Liu.
\newblock Predicting drug-disease associations by using similarity constrained
  matrix factorization.
\newblock {\em BMC bioinformatics}, 19:1--12, 2018.

\bibitem{li2021scmfmda}
Lei Li, Zhen Gao, Yu-Tian Wang, Ming-Wen Zhang, Jian-Cheng Ni, Chun-Hou Zheng,
  and Yansen Su.
\newblock Scmfmda: predicting microrna-disease associations based on similarity
  constrained matrix factorization.
\newblock {\em PLoS computational biology}, 17(7):e1009165, 2021.

\bibitem{zhao2015nonconvex}
Tuo Zhao, Zhaoran Wang, and Han Liu.
\newblock Nonconvex low rank matrix factorization via inexact first order
  oracle.
\newblock {\em Advances in Neural Information Processing Systems},
  458:461--462, 2015.

\bibitem{ge2016matrix}
Rong Ge, Jason~D Lee, and Tengyu Ma.
\newblock Matrix completion has no spurious local minimum.
\newblock {\em Advances in neural information processing systems}, 29, 2016.

\bibitem{ge2017no}
Rong Ge, Chi Jin, and Yi~Zheng.
\newblock No spurious local minima in nonconvex low rank problems: A unified
  geometric analysis.
\newblock In {\em International conference on machine learning}, pages
  1233--1242. PMLR, 2017.

\bibitem{li2019non}
Qiuwei Li, Zhihui Zhu, and Gongguo Tang.
\newblock The non-convex geometry of low-rank matrix optimization.
\newblock {\em Information and Inference: A Journal of the IMA}, 8(1):51--96,
  2019.

\bibitem{koren2009bellkor}
Yehuda Koren.
\newblock The bellkor solution to the netflix grand prize.
\newblock {\em Netflix prize documentation}, 81(2009):1--10, 2009.

\bibitem{koren2009matrix}
Yehuda Koren, Robert Bell, and Chris Volinsky.
\newblock Matrix factorization techniques for recommender systems.
\newblock {\em Computer}, 42(8):30--37, 2009.

\bibitem{donoho2023screenot}
David Donoho, Matan Gavish, and Elad Romanov.
\newblock Screenot: Exact mse-optimal singular value thresholding in correlated
  noise.
\newblock {\em The Annals of Statistics}, 51(1):122--148, 2023.

\bibitem{nagai2017overview}
Akiko Nagai, Makoto Hirata, Yoichiro Kamatani, Kaori Muto, Koichi Matsuda,
  Yutaka Kiyohara, Toshiharu Ninomiya, Akiko Tamakoshi, Zentaro Yamagata,
  Taisei Mushiroda, et~al.
\newblock Overview of the biobank japan project: Study design and profile.
\newblock {\em Journal of epidemiology}, 27(Supplement\_III):S2--S8, 2017.

\bibitem{bycroft2018uk}
Clare Bycroft, Colin Freeman, Desislava Petkova, Gavin Band, Lloyd~T Elliott,
  Kevin Sharp, Allan Motyer, Damjan Vukcevic, Olivier Delaneau, Jared
  O’Connell, et~al.
\newblock The uk biobank resource with deep phenotyping and genomic data.
\newblock {\em Nature}, 562(7726):203--209, 2018.

\bibitem{kurki2023finngen}
Mitja~I Kurki, Juha Karjalainen, Priit Palta, Timo~P Sipil{\"a}, Kati
  Kristiansson, Kati~M Donner, Mary~P Reeve, Hannele Laivuori, Mervi Aavikko,
  Mari~A Kaunisto, et~al.
\newblock Finngen provides genetic insights from a well-phenotyped isolated
  population.
\newblock {\em Nature}, 613(7944):508--518, 2023.

\bibitem{kaiser1958varimax}
Henry~F Kaiser.
\newblock The varimax criterion for analytic rotation in factor analysis.
\newblock {\em Psychometrika}, 23(3):187--200, 1958.

\bibitem{mazumder2010spectral}
Rahul Mazumder, Trevor Hastie, and Robert Tibshirani.
\newblock Spectral regularization algorithms for learning large incomplete
  matrices.
\newblock {\em The Journal of Machine Learning Research}, 11:2287--2322, 2010.

\bibitem{softImpute}
Trevor Hastie and Rahul Mazumder.
\newblock {\em softImpute: Matrix Completion via Iterative Soft-Thresholded
  SVD}, 2021.
\newblock R package version 1.4-1.

\bibitem{zhang2022heteroskedastic}
Anru~R Zhang, T~Tony Cai, and Yihong Wu.
\newblock Heteroskedastic pca: Algorithm, optimality, and applications.
\newblock {\em The Annals of Statistics}, 50(1):53--80, 2022.

\bibitem{landa2025dyson}
Boris Landa and Yuval Kluger.
\newblock The dyson equalizer: Adaptive noise stabilization for low-rank signal
  detection and recovery.
\newblock {\em Information and Inference: A Journal of the IMA}, 14(1):iaae036,
  2025.

\end{thebibliography}


\begin{thebibliography}{1}

\bibitem{sakaue2021cross}
Saori Sakaue, Masahiro Kanai, Yosuke Tanigawa, Juha Karjalainen, Mitja Kurki,
  Seizo Koshiba, Akira Narita, Takahiro Konuma, Kenichi Yamamoto, Masato
  Akiyama, et~al.
\newblock A cross-population atlas of genetic associations for 220 human
  phenotypes.
\newblock {\em Nature genetics}, 53(10):1415--1424, 2021.

\bibitem{10.1093/nar/gkz836}
Susan Fairley, Ernesto Lowy-Gallego, Emily Perry, and Paul Flicek.
\newblock The international genome sample resource (igsr) collection of open
  human genomic variation resources.
\newblock {\em Nucleic Acids Research}, 48(D1):D941--D947, 10 2019.

\bibitem{purcell2007plink}
Shaun Purcell, Benjamin Neale, Kathe Todd-Brown, Lori Thomas, Manuel~AR
  Ferreira, David Bender, Julian Maller, Pamela Sklar, Paul~IW De~Bakker,
  Mark~J Daly, et~al.
\newblock Plink: a tool set for whole-genome association and population-based
  linkage analyses.
\newblock {\em The American journal of human genetics}, 81(3):559--575, 2007.

\bibitem{10002015global}
{The 1000 Genomes Project Consortium}.
\newblock A global reference for human genetic variation.
\newblock {\em Nature}, 526(7571):68--74, 2015.

\bibitem{fairley2020international}
Susan Fairley, Ernesto Lowy-Gallego, Emily Perry, and Paul Flicek.
\newblock {The International Genome Sample Resource (IGSR)} collection of open
  human genomic variation resources.
\newblock {\em Nucleic acids research}, 48(D1):D941--D947, 2020.

\end{thebibliography}

\end{document}


\maketitle
\tableofcontents

\newpage

\section{Extensions for LEARNER}

In this section, we present a few methodological extensions for LEARNER.

\subsection{Column- and row-space specific penalties}

When presenting LEARNER in the main text, we used the same penalty (i.e., $\lambda_1$) for the difference between the latent row and column spaces between the source and target populations. Here, we describe how LEARNER can be adapted for having column- and row-space specific penalties. While more computationally expensive, this approach is better suited for settings where the degrees of similarity in the latent row and column spaces between the source and target populations are believed to be highly different (e.g., only the latent row spaces are similar between the source and target populations). 

In this case, the LEARNER estimator of $\Theta_0$ is given by $\bar{U}\bar{V}^{\top}$ where $(\bar{U}, \bar{V})$ is the solution to 
\begin{equation} \label{eq: obj lat dif penalties}
        \argmin_{U \in \mathbb{R}^{p \times r}, V \in \mathbb{R}^{q \times r}} \big\{ \| U V^{\top} - Y_0 \|_F^2 + \lambda_{1,1}\| \mathcal{P}_{\perp}(\hat{U}_{1})U \|_F^2 + \lambda_{1,2}\|  \mathcal{P}_{\perp}(\hat{V}_{1})V \|_F^2  + \lambda_2 \| U^{\top} U - V^{\top} V \|_F^2 \big\}.
\end{equation}
When $\lambda_{1,1} = \lambda_{1,2}$, this reduces to the LEARNER estimator in the main text. Letting $\overline{f}$ denote the objective function in (\ref{eq: obj lat dif penalties}), it is straightforward to see that the gradients are given by
\begin{align*}
    \nabla_{U} \overline{f}(U,V) & = 2(UV^{\top}V - Y_0V) + 2 \lambda_{1,1} \mathcal{P}_{\perp}(\hat{U}_{1})U + 4 \lambda_2 U(U^{\top}U - V^{\top}V) \\
    \nabla_{V} \overline{f}(U,V) & = 2(VU^{\top}U - Y_0^{\top}U) + 2 \lambda_{1,2} \mathcal{P}_{\perp}(\hat{V}_{1})V + 4 \lambda_2 V(V^{\top}V - U^{\top}U).
\end{align*}
The same numerical optimization approach presented in Algorithm 1 of the main text can be applied with this new objective function and gradients. Cross-validation can again be used to select the degree of transfer learning, with the only change being that the search is conducted over a grid for $(\lambda_{1,1}, \lambda_{1,2}, \lambda_2)$ rather than over $(\lambda_1, \lambda_2)$.

\subsection{Selecting the penalties based on an external dataset} \label{sec: external}

Recall that we considered using cross-validation to select the penalty parameters $(\lambda_1, \lambda_2)$ for LEARNER in the main text. However, this approach can perform sub-optimally in the presence of correlated noise. Here, we consider an alternative approach that is based on using an external dataset rather than held-out entries of $Y_0$ to estimate the MSE of LEARNER for each candidate $(\lambda_1, \lambda_2)$. 

Suppose that we observe $Y_0^{\mathrm{ext}} \in \mathbb{R}^{p \times q}$ in an external dataset such that
\begin{equation}
    Y_0^{\mathrm{ext}} = \Theta_0 + Z_0^{\mathrm{ext}} \label{eq: Y_0 ext}
\end{equation}
where $Z_0^{\mathrm{ext}}$ follows the same distribution as $Z_0$. The matrix $Y_0^{\mathrm{ext}}$ may have $p^{\prime}$ missing rows and $q^{\prime}$ missing columns due to inconsistencies between the measured variables in the data sources (e.g., the external source may not collect data on some of the phenotypes and genotypes that were included in $Y_0$ and $Y_1$). For each candidate $(\lambda_1, \lambda_2)$ pair, we apply LEARNER (based on $Y_0$ and $Y_1$) and evaluate its MSE using the external dataset, i.e., 
\begin{equation*}
    \text{MSE}(\lambda_1, \lambda_2) = \frac{1}{|\Omega_{Y_0^{\mathrm{ext}}}|} \sum_{(i,j) \in \Omega_{Y_0^{\mathrm{ext}}}} (\hat{\Theta}^{\mathrm{LEARNER},\lambda_1,\lambda_2}_{ij} - Y_{0,ij}^{\mathrm{ext}})^2
\end{equation*}
where $\hat{\Theta}^{\mathrm{LEARNER},\lambda_1,\lambda_2}$ denotes the LEARNER estimate with $(\lambda_1, \lambda_2)$ and $\Omega_{Y_0^{\mathrm{ext}}}$ denotes the set of non-missing entries in $Y_0^{\mathrm{ext}}$. We select the value of $(\lambda_1, \lambda_2)$ with the smallest MSE.

\section{Additional details on the simulation study} \label{appendix: simulations}

\subsection{Hyperparameter settings}

Recall that LEARNER involves specifying candidate $\lambda_1, \lambda_2$ values as well as several hyperparameters for the numerical optimization algorithm (i.e., initial values, step size, maximum number of iterations, tolerance for $|\epsilon_t -\epsilon_{t-1}|$). In this subsection, we describe the hyperparameter values used in the simulations.

In all simulation scenarios, we initialized $U$ by $\hat{U}_1 \hat{\Lambda}_1^{1/2}$ and initialized $V^{\top}$ by $\hat{\Lambda}_1^{1/2} \hat{V}_1^{\top}$. We set the tolerance for $|\epsilon_t -\epsilon_{t-1}|$ to be 0.001 and set the maximum number of iterations to 75. We considered 5 candidate $\lambda_1$ values in an equally spaced grid on the log scale (lower and upper bounds are described below). Similarly, we considered 5 candidate $\lambda_2$ values in an equally spaced grid on the log-scale, resulting in 25 candidate values for $(\lambda_1, \lambda_2)$.

The step size and grid of values for $\lambda_1, \lambda_2$ depended on the simulation scenarios. In general, we set the step size to a value where the optimization algorithm clearly converged in most iterations, and we set the upper and lower bounds for $\lambda_1$ and $\lambda_2$ to values so that the selected $(\lambda_1, \lambda_2)$ fell well within the boundary of the grid in most iterations. Specifically, in the independent noise simulations, we set these parameters as follows:
    \begin{itemize}
        \item High similarity scenarios: We used a step size of $c=0.0035$. The candidate $\lambda_1$ and $\lambda_2$ values ranged from $10^{-4}$ to $10^{4}$.
        \item Moderate similarity scenarios: We used a step size of $c=0.035$. In the rectangular matrix setting, the candidate $\lambda_1$ values ranged from $10^{0}$ to $10^{4}$, and the candidate $\lambda_2$ values ranged from $10^{-2}$ to $10^{1}$. In the square matrix setting, the candidate $\lambda_1$ values ranged from $10^{1}$ to $10^{3}$, and the candidate $\lambda_2$ values ranged from $10^{-6}$ to $10^{0}$.
        \item Low similarity scenarios: We used a step size of $c=0.07$. The candidate $\lambda_1$ values ranged from $10^{0}$ to $10^{4}$, and the candidate $\lambda_2$ values ranged from $10^{-2}$ to $10^{1}$.
    \end{itemize}
In the correlated noise simulations, we used a step size of $c=0.035$. The candidate $\lambda_1$ and $\lambda_2$ values ranged from $10^{-4}$ to $10^{4}$.

Recall that LEARNER uses ScreeNOT for rank selection. ScreeNOT requires specifying one hyperparameter: a loose upper bound on the rank $r$. We set this value to $\lfloor \min\{50, 5000\} / 3 \rfloor = 16$ in all simulation scenarios.

\subsection{Results in the square matrix settings}

\begin{figure} [H]
    \centering
    \includegraphics[width=0.75\textwidth]{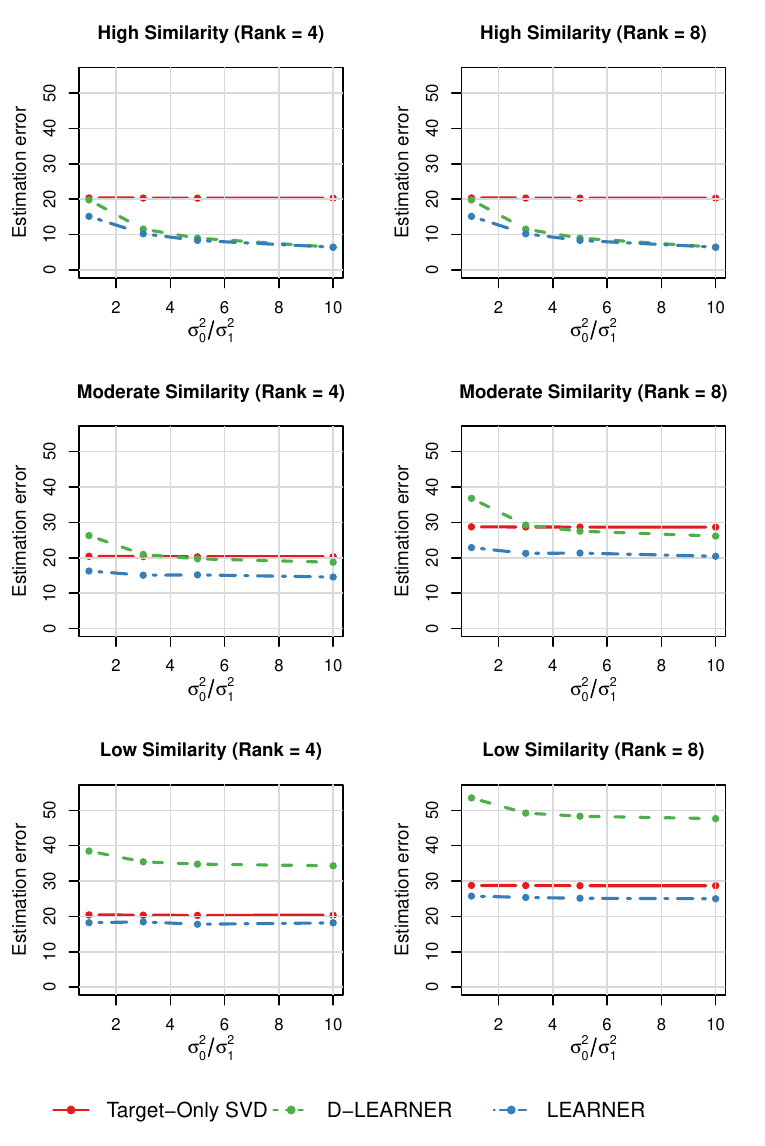}
    \caption{Simulation results in the square matrix settings with independent noise.\label{fig: independent noise square}}
\end{figure}

Figure \ref{fig: independent noise square} summarizes the simulation results in the square matrix settings with independent noise. The same trends held as in the rectangular matrix settings with independent noise presented in the main text.

\subsection{Simulations with an external dataset for selecting the LEARNER penalties}

We present simulations that evaluate the performance of LEARNER when using an external dataset to select the penalties (i.e., as described in Section \ref{sec: external}). Specifically, we apply this version of LEARNER to the simulation study described in Section 3.2 of the main text (i.e., with correlated noise). We generated $Y_0^{\mathrm{ext}}$ by equation (\ref{eq: Y_0 ext}). We set these parameters as follows:
    \begin{itemize}
        \item High similarity scenarios: We used a step size of $c=0.0075$ and a maximum number of iterations of 200. The candidate $\lambda_1$ values ranged from $10^{0}$ to $10^{4}$, and the candidate $\lambda_2$ values ranged from $10^{-4}$ to $10^{4}$.
        \item Moderate similarity scenarios: We used a step size of $c=0.05$ and a maximum number of iterations of 75. The candidate $\lambda_1$ values ranged from $10^{0}$ to $10^{4}$, and the candidate $\lambda_2$ values ranged from $10^{-2}$ to $10^{2}$.
        \item Low similarity scenarios: We used a step size of $c=0.07$ and a maximum number of iterations of 75. The candidate $\lambda_1$ values ranged from $10^{0}$ to $10^{4}$, and the candidate $\lambda_2$ values ranged from $10^{-2}$ to $10^{2}$.
    \end{itemize}

\begin{figure} [H]
    \centering
    \includegraphics[width=\textwidth]{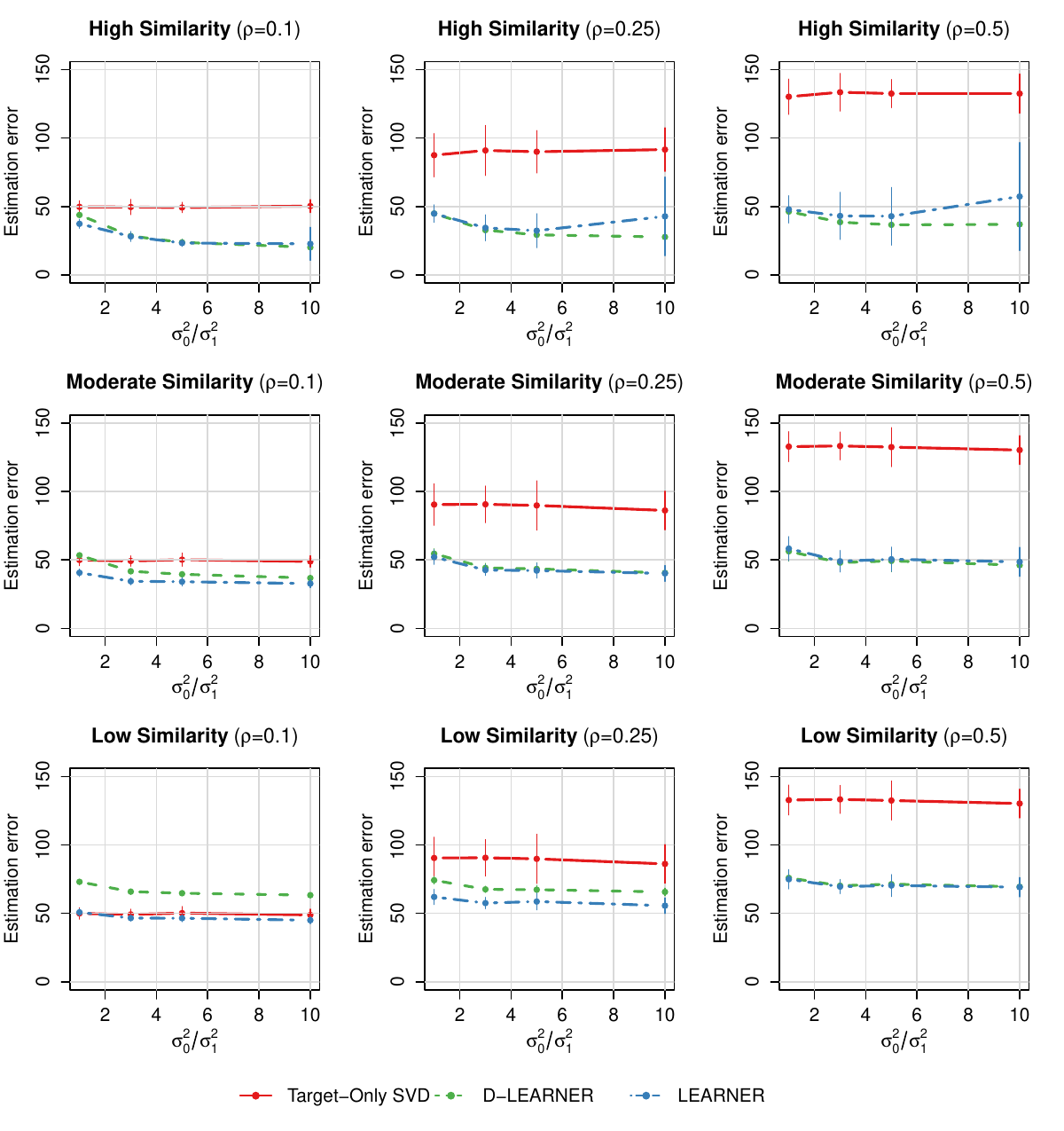}
    \caption{Simulation results under the correlated noise settings, where LEARNER used an external dataset to select $(\lambda_1, \lambda_2)$. The error bars correspond to the standard deviation of the estimation error across the 50 repetitions. \label{fig: correlated noise ext}}
\end{figure}

The simulation results are summarized in Figure \ref{fig: correlated noise ext}. LEARNER generally outperformed or performed comparable to the target-only SVD and D-LEARNER methods. As the degree of correlation increased, the discrepancy between the LEARNER and D-LEARNER methods generally decreased.  

\section{Additional details on the data application} \label{appendix: application}

\subsection{Data processing} \label{sec: data processing}

In this section, we describe the additional data processing steps that were performed after obtaining the GWAS summary statistics from \cite{sakaue2021cross}. 

We performed a variant screening procedure similar to that described by \cite{sakaue2021cross}. We considered estimates insignificant when either the p-value for a standard Wald test was greater than 0.001 or the standard error was greater than 0.2. We excluded variants that had all insignificant associations across the 145 phenotypes in either the BBJ or European populations. We also used The International Genome Sample Resource (IGSR) \cite{10.1093/nar/gkz836} --- a fully open human genomic resource built upon 1000 Genomes Project, as the Minor Allele Frequency (MAF) standardization reference. As a result, we further excluded variants whose MAF is less than 0.005 for either Eastern Asian or European population. A total of 208,583 variants were left after this screening step. We then performed linkage disequilibrium (LD) pruning using PLINK \cite{purcell2007plink} ($‘$--indep-pairwise 50 5 0.1$’$) with an LD reference from the 1000 Genomes Project phase 3 data \cite{10002015global, fairley2020international}. This resulted in 27,696 remaining genetic variants. We then removed 572 of these genetic variants because they had missing genetic association values across all of the 27 phenotypes that were only measured in the UKB data and not the FinnGen data. We then removed variants located in the major histocompatibility complex region (chromosome 6: 25–34 megabase). The final analytic datasets contained 25,415 genetic variants and 145 phenotypes, where there were 0.006\% missing entries in $Y_0$ and 0.102\% missing entries in $Y_1$. These missing entries were set to 0 for all methods for consistency. 

After performing the variant screening, we standardized the genetic associations using the IGSR dataset. If the MAF of a variant in a given population is $p$, then the MAF standardization coefficient is $\sqrt{p(1-p)}$. Among the 4978 individuals in the IGSR datasets, 858 are of Eastern and Southern Eastern Asian ancestry, and 670 are of European ancestry. We calculated the MAF values based on these individuals.

\subsection{Exploratory data analyses} \label{sec: exploratory appendix}

\begin{figure}[H]
    \centering
        \includegraphics[width=0.5\textwidth]{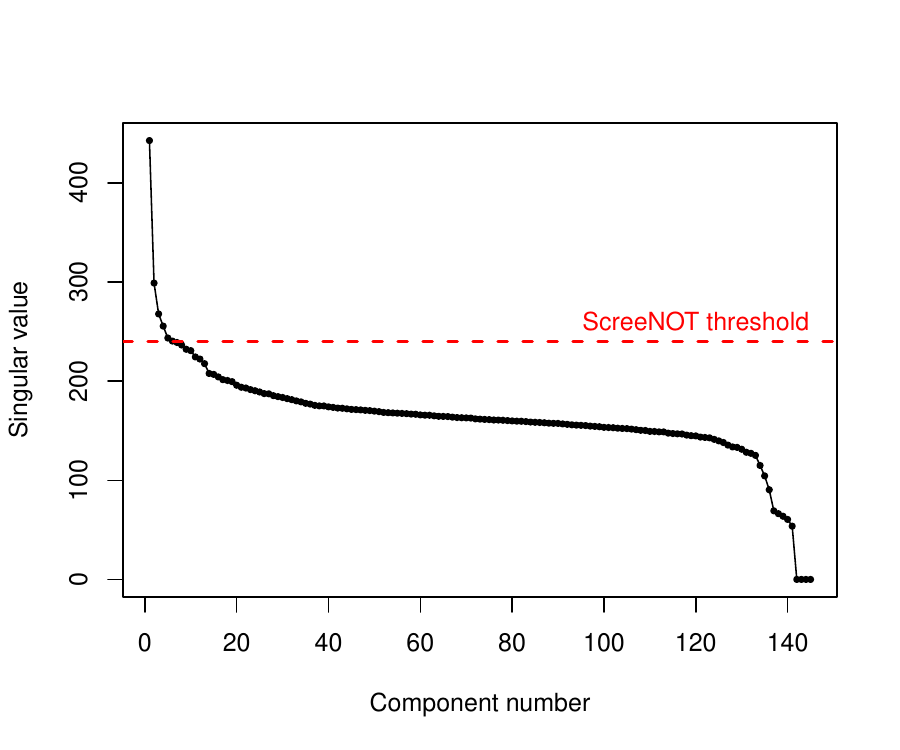}
        \caption{Scree plot based on the source population data. \label{fig: scree}}
\end{figure}

\begin{figure}[H]
    \centering
    \begin{subfigure}{0.5\textwidth}
        \centering
        \includegraphics[width=\textwidth]{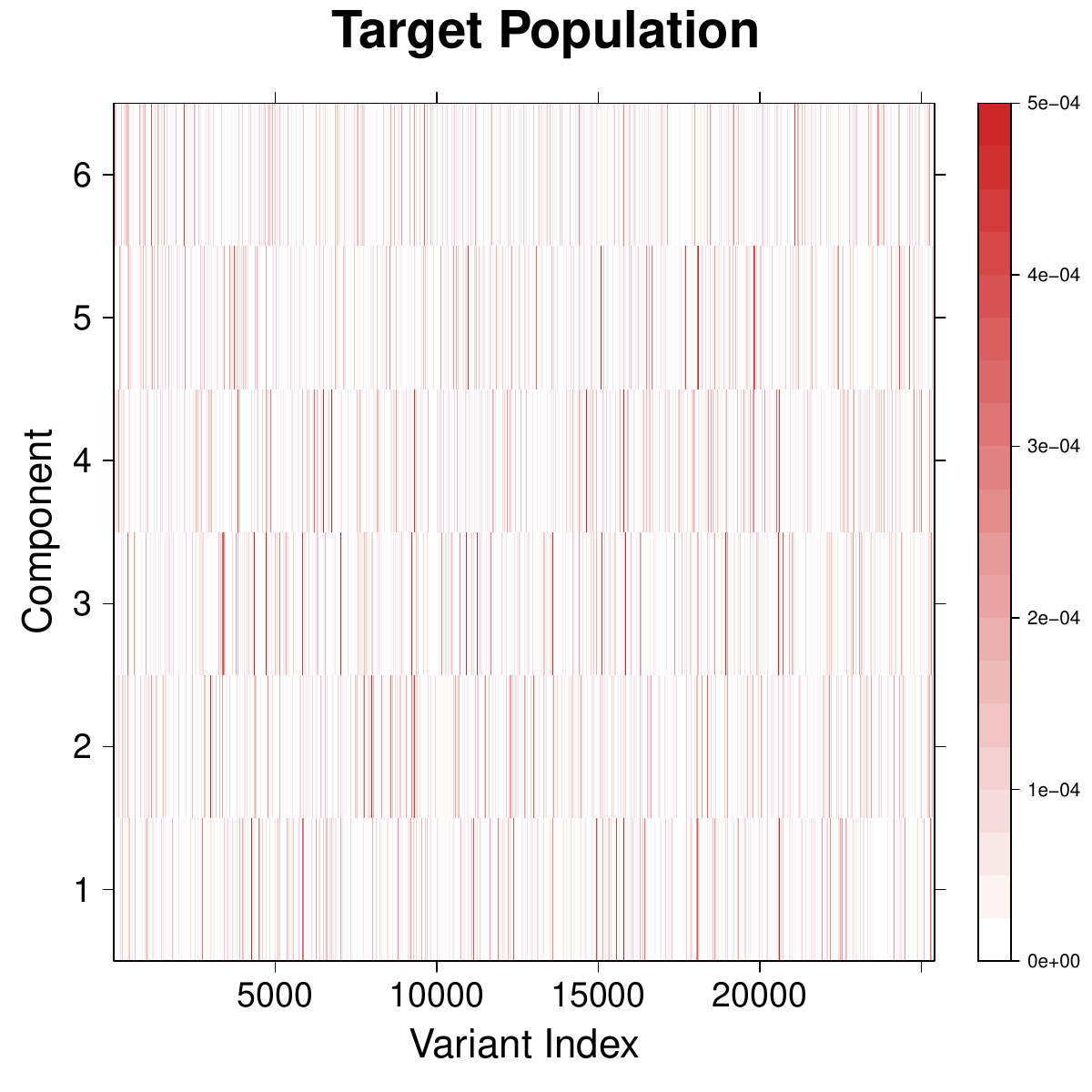}
    \end{subfigure}%
    ~ 
    \begin{subfigure}{0.5\textwidth}
        \centering
        \includegraphics[width=\textwidth]{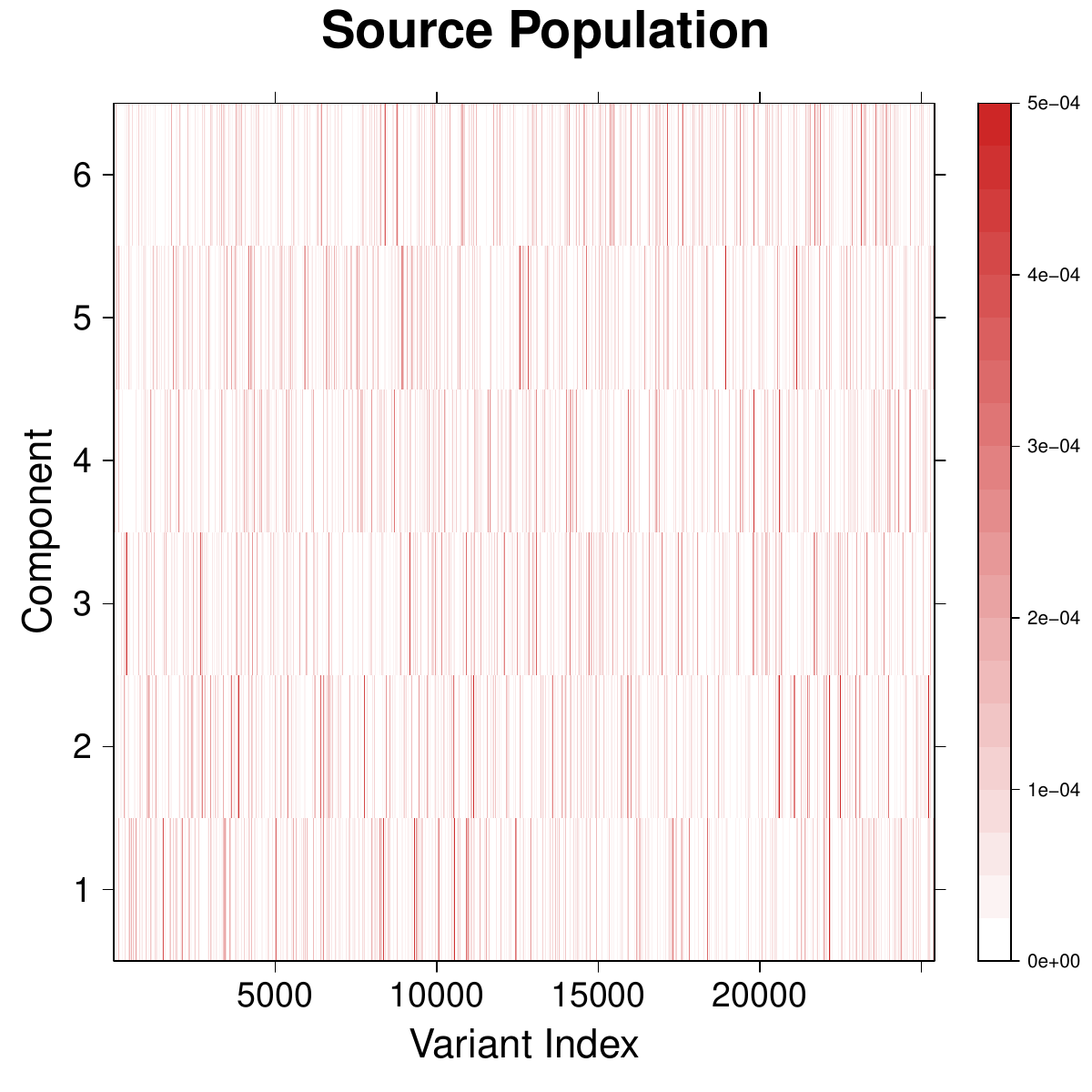}
    \end{subfigure}
    \caption{Heatmap of the matrix of variant contribution scores in the target population (left panel) and source population (right panel). The variants are ordered based on their chromosome and position number. \label{fig: variant contribution}}
\end{figure}

\begin{figure}[H]
    \centering
    \begin{subfigure}{0.5\textwidth}
        \centering
        \includegraphics[width=\textwidth]{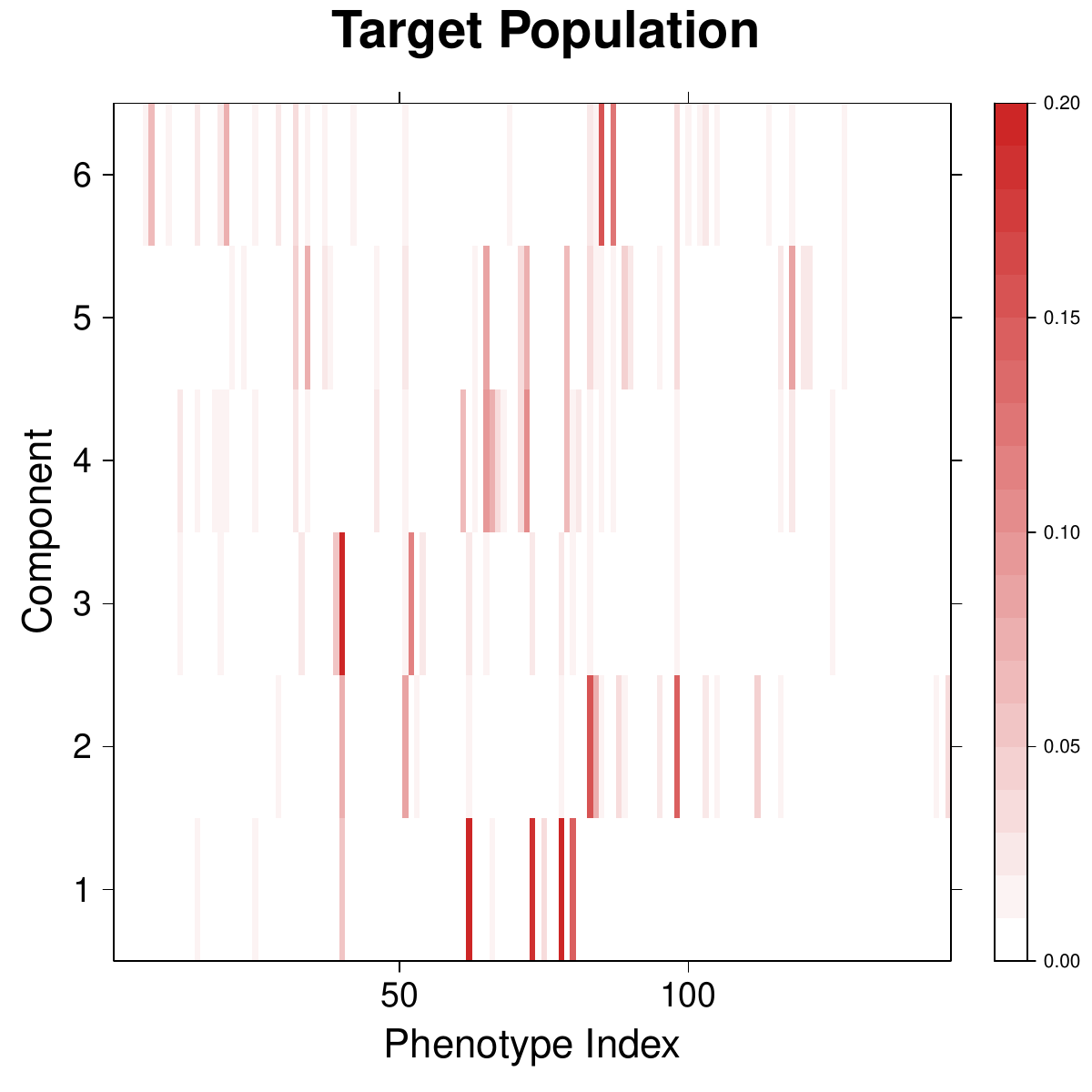}
    \end{subfigure}%
    ~ 
    \begin{subfigure}{0.5\textwidth}
        \centering
        \includegraphics[width=\textwidth]{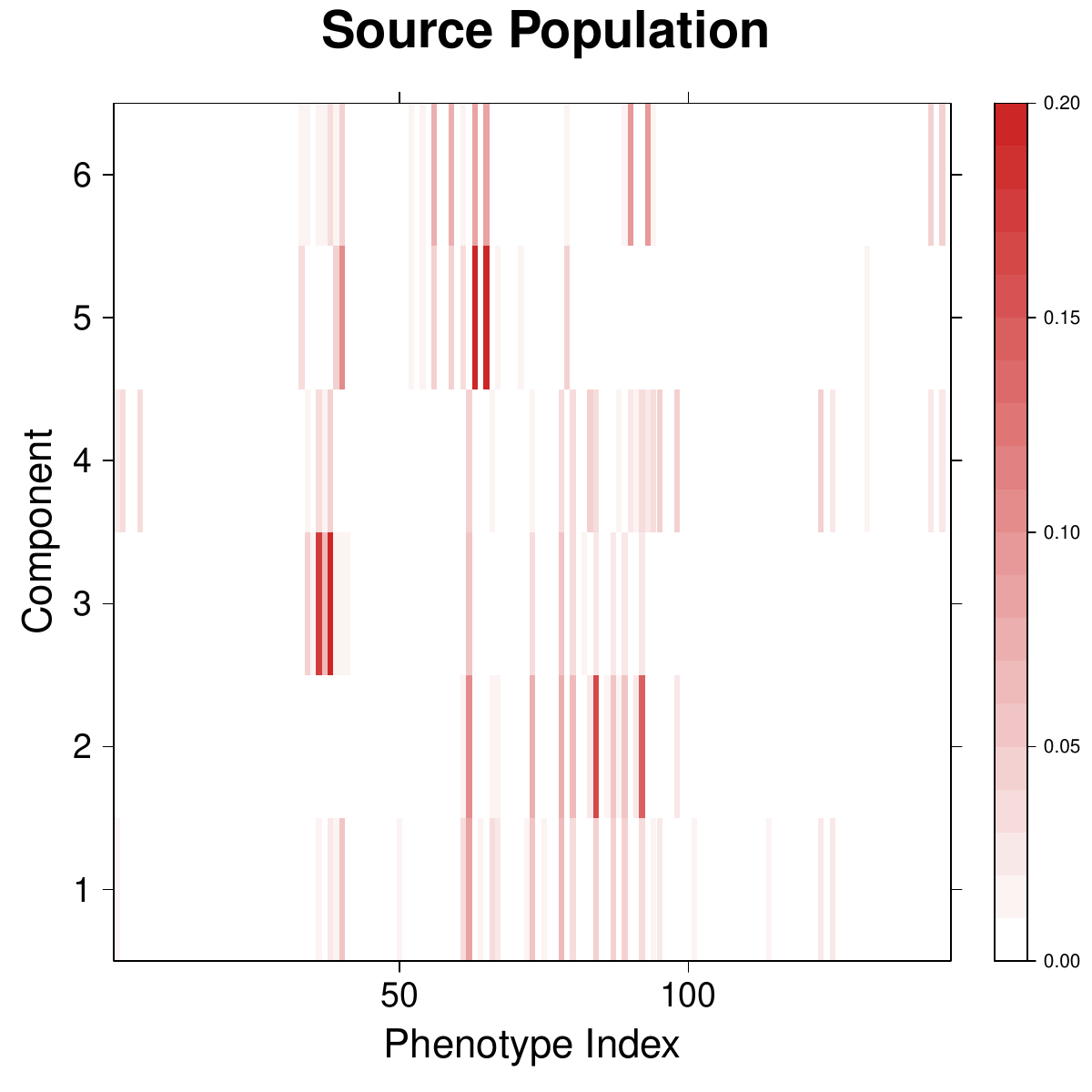}
    \end{subfigure}
    \caption{Heatmap of the matrix of phenotype contribution scores in the target population (left panel) and source population (right panel). The phenotypes are ordered based on their ICD-10 category.\label{fig: phenotype contribution}}
\end{figure}

\begin{table}[H]
\caption{Top phenotypes in each latent factor in the source population. For each latent factor, we list the four phenotypes with the highest contribution scores. The phenotype contribution scores are in parentheses. The phenotypes related to the latent factor characterization are in bold text.\label{tab: key phenotypes source}}
\begin{center}
\begin{tabular}{p{0.06\textwidth}p{0.25\textwidth}p{0.60\textwidth}} \hline 
            Factor & Characterization & Top phenotypes   \\ \hline
            1 & Angina & \textbf{Angina pectoris (0.08)}, \textbf{stable angina pectoris (0.06)}, myocardial infarction (0.06), type 2 diabetes (0.05) \\
            2 & Asthma & \textbf{Asthma (0.17)}, \textbf{pediatric asthma (0.15)}, angina pectoris (0.10), myocardial infarction (0.07) \\
            3 & Thyroid disorders & \textbf{Hypothyroidism (0.21)}, \textbf{Hashimoto's disease (0.17)}, \textbf{hyperthyroidism (0.06)}, stable angina pectoris (0.06)  \\
            4 & Unclear & Hypothyroidism (0.05), angina pectoris (0.05), allergic rhinitis (0.05), pneumonia (0.04) \\ 
            5 & Aneurysm & \textbf{Unruptured cerebral aneurysm (0.22)}, \textbf{cerebral aneurysm (0.22)}, type 2 diabetes (0.10), \textbf{subarachnoid hemorrhage (0.05)} \\
            6 & Pulmonary Disorders & \textbf{Interstitial lung disease (0.09)}, \textbf{pulmonary fibrosis (0.09)}, unruptured cerebral aneurysm (0.09), cerebral aneurysm (0.09) \\ \hline
\end{tabular}
\end{center}
\end{table} 

\begin{table}[H]
\caption{Top phenotypes in each latent factor in the target population. For each latent factor, we list the four phenotypes with the highest contribution scores. The phenotype contribution scores are in parentheses. The phenotypes related to the latent factor characterization are in bold text. \label{tab: key phenotypes target}}
\begin{center}
\begin{tabular}{p{0.06\textwidth}p{0.25\textwidth}p{0.60\textwidth}} \hline 
            Factor & Characterization & Top phenotypes   \\ \hline
            1 & Angina & \textbf{Angina pectoris (0.08)}, \textbf{stable angina pectoris (0.06)}, myocardial infarction (0.06), type 2 diabetes (0.05) \\
            2 & Asthma & \textbf{Asthma (0.17)}, \textbf{pediatric asthma (0.15)}, angina pectoris (0.10), myocardial infarction (0.07) \\
            3 & Thyroid disorders & \textbf{Hypothyroidism (0.21)}, \textbf{Hashimoto's disease (0.17)}, \textbf{hyperthyroidism (0.06)}, stable angina pectoris (0.06)  \\
            4 & Unclear & Hypothyroidism (0.05), angina pectoris (0.05), allergic rhinitis (0.05), pneumonia (0.04) \\ 
            5 & Aneurysm & \textbf{Unruptured cerebral aneurysm (0.22)}, \textbf{cerebral aneurysm (0.22)}, type 2 diabetes (0.10), \textbf{subarachnoid hemorrhage (0.05)} \\
            6 & Pulmonary Disorders & \textbf{Interstitial lung disease (0.09)}, \textbf{pulmonary Fibrosis (0.09)}, unruptured cerebral aneurysm (0.09), cerebral aneurysm (0.09) \\ \hline
\end{tabular}
\end{center}
\end{table}

\subsection{LEARNER and D-LEARNER illustration} \label{sec: lat application appendix}

\subsubsection{Hyperparameter settings and selection for LEARNER} \label{sec: hyperparameter application}

We initialized $U$ and $V$ in the same manner as described in Appendix \ref{appendix: simulations}. We also used the same tolerance for $|\epsilon_t -\epsilon_{t-1}|$. We considered 10 candidate values for $\lambda_1$ on an equally spaced grid on the log-scale, ranging from $10^1$ to $10^{4}$. Similarly, we considered 10 candidate values for $\lambda_2$ on an equally spaced a grid on the log-scale, ranging from and $10^{-2}$ and $10^{1}$. We used a step size of 0.04 and a maximum number of iterations of 100. 

\begin{figure}[H]
    \centering
    \includegraphics[width=0.5\textwidth]{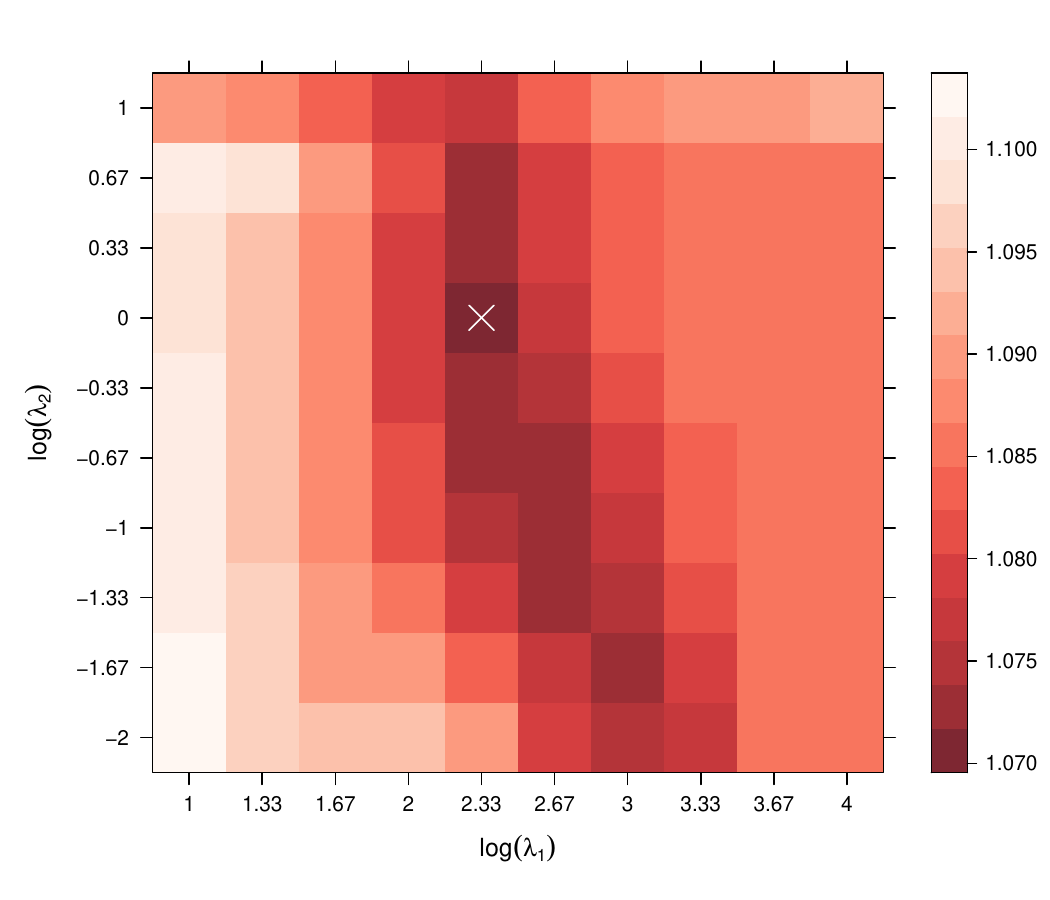}
    \caption{Heatmap of the held-out MSE for each candidate $(\lambda_1, \lambda_2)$. The white X mark indicates the candidate $(\lambda_1, \lambda_2)$ with the smallest held-out MSE. \label{fig: nuisance}}
\end{figure}

\begin{figure} [H]
    \centering
    \includegraphics[width=0.6\textwidth]{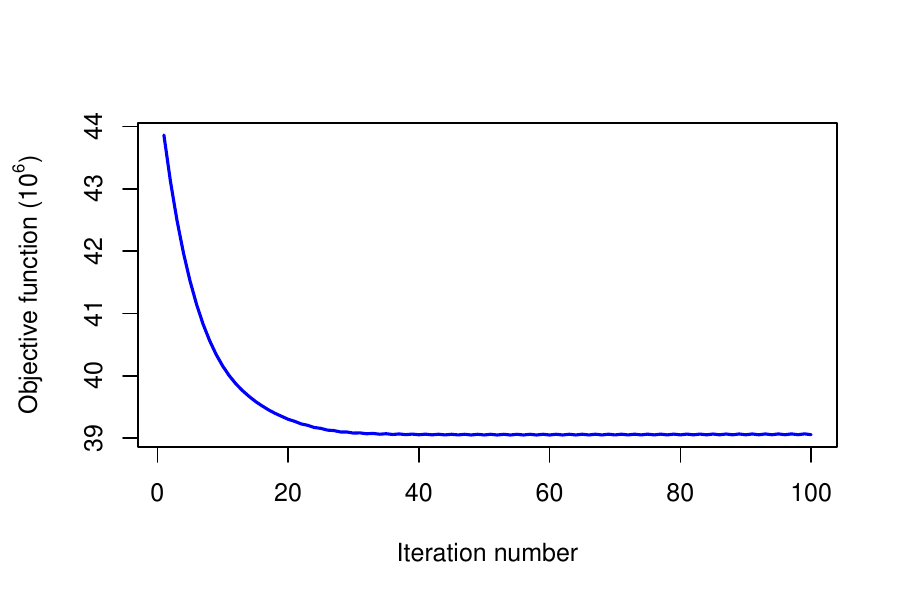}
    \caption{Convergence of the numerical optimization algorithm for solving the LEARNER objective function based on the full matrix $Y_0$.\label{fig: objfunc}}
\end{figure}

Figure \ref{fig: nuisance} shows the held-out MSE for each candidate $(\lambda_1, \lambda_2)$. The $(\lambda_1, \lambda_2)$ attaining the smallest MSE was near the center of the grid. Using $(\lambda_1^{(\mathrm{best})}, \lambda_2^{(\mathrm{best})})=(10^{2.33}, 10^0)$, the convergence of the numerical optimization algorithm for solving the LEARNER objective function based on the full matrix $Y_0$ (i.e., the last line in Algorithm 2) is illustrated in Figure \ref{fig: objfunc}.

\subsubsection{Results from LEARNER and D-LEARNER}

\begin{figure}[H]
    \centering
    \begin{subfigure}[t]{0.4\textwidth}
        \includegraphics[width=\linewidth]{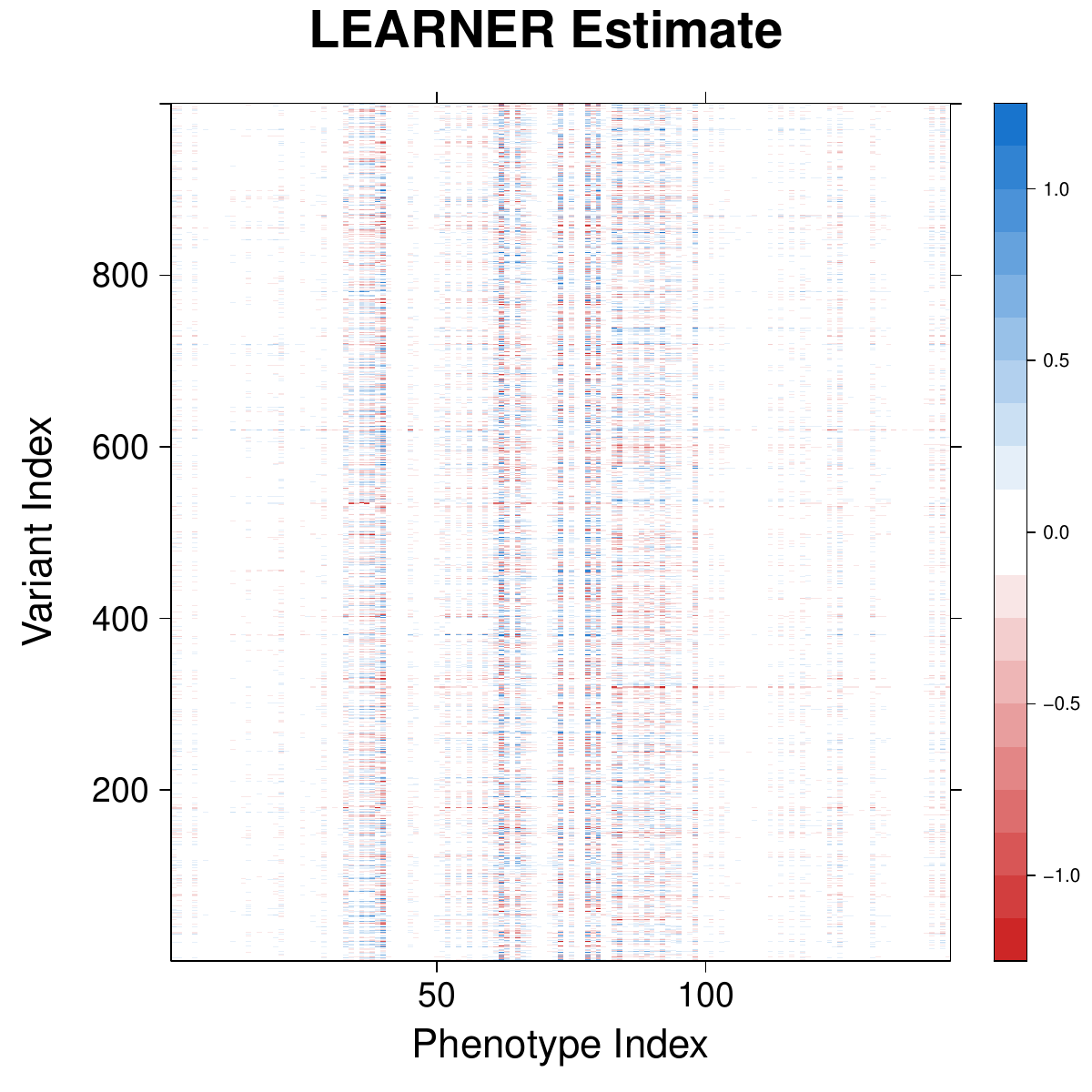}
    \end{subfigure}%
    \begin{subfigure}[t]{0.4\textwidth}
        \includegraphics[width=\linewidth]{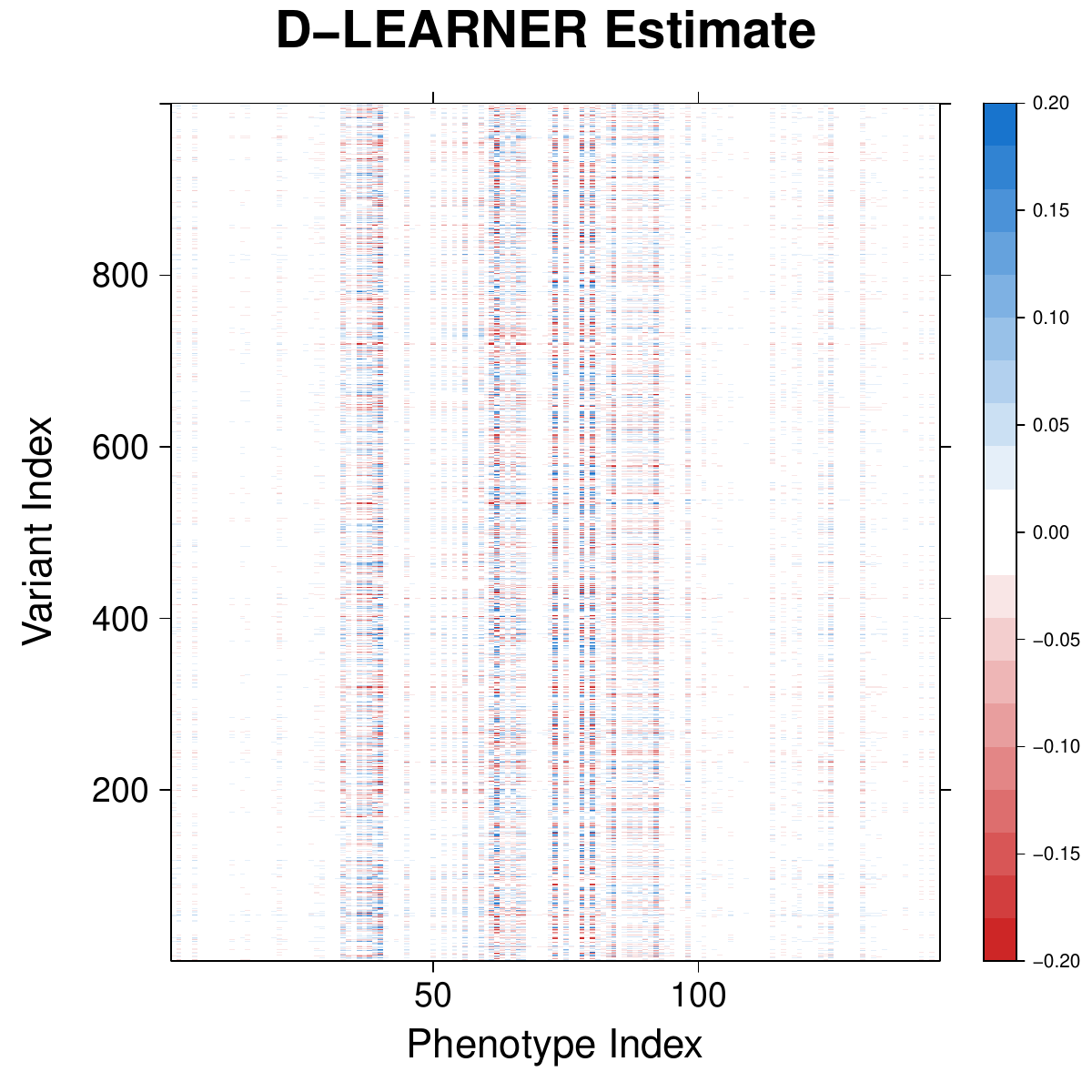} 
    \end{subfigure}
    \caption{Heatmap of the LEARNER and D-LEARNER estimates of $\Theta_0$. A random subset of 1000 genetic variants (out of the 25,415) and all phenotypes are illustrated. The phenotypes are ordered based on their ICD-10 category, and the variants are ordered based on their chromosome and position number.}
    \label{fig:learner dlearner estimates}
\end{figure}

\begin{figure}[H]
    \centering
    \begin{subfigure}{0.4\textwidth}
        \centering
        \includegraphics[width=\textwidth]{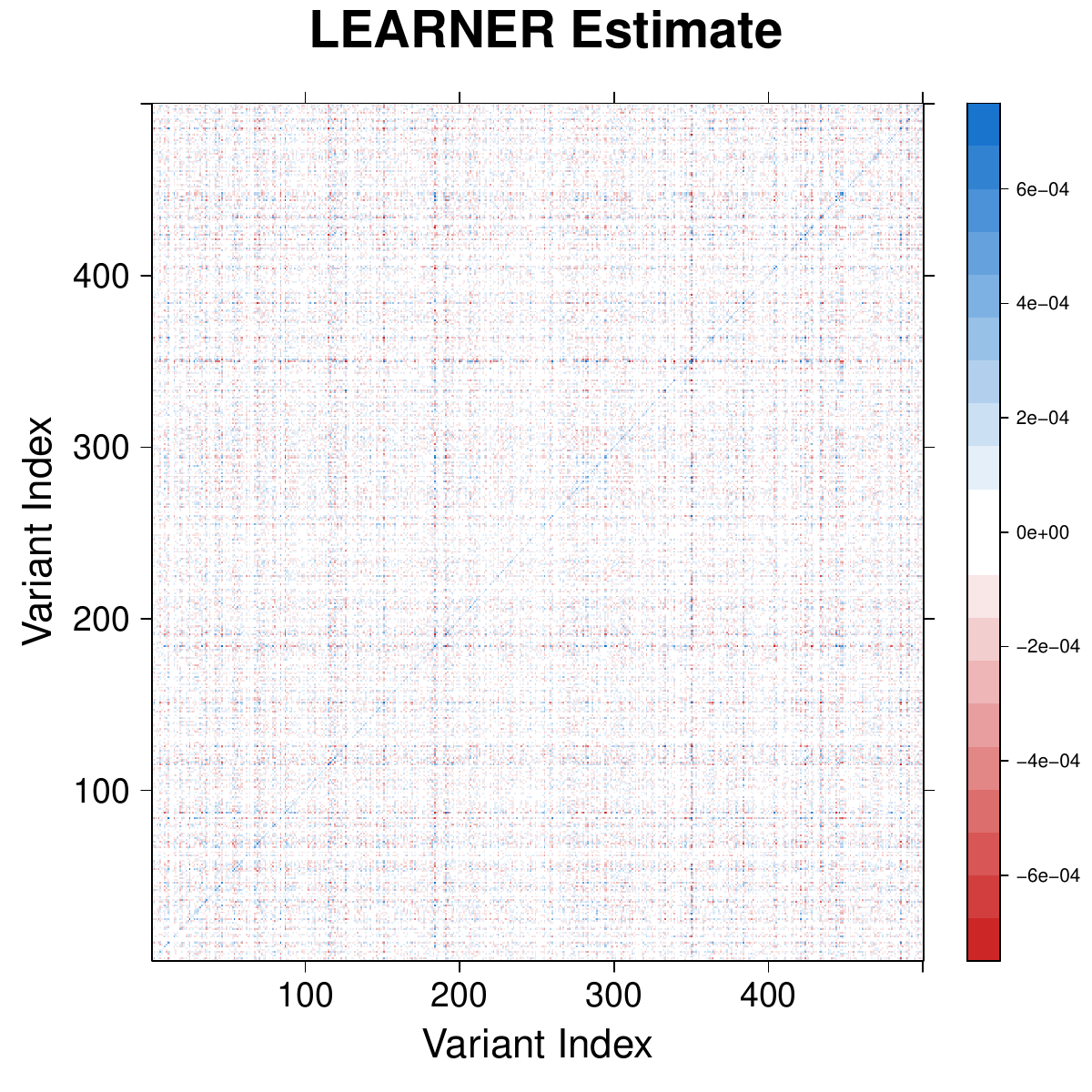}
    \end{subfigure}%
    ~ 
    \begin{subfigure}{0.4\textwidth}
        \centering
        \includegraphics[width=\textwidth]{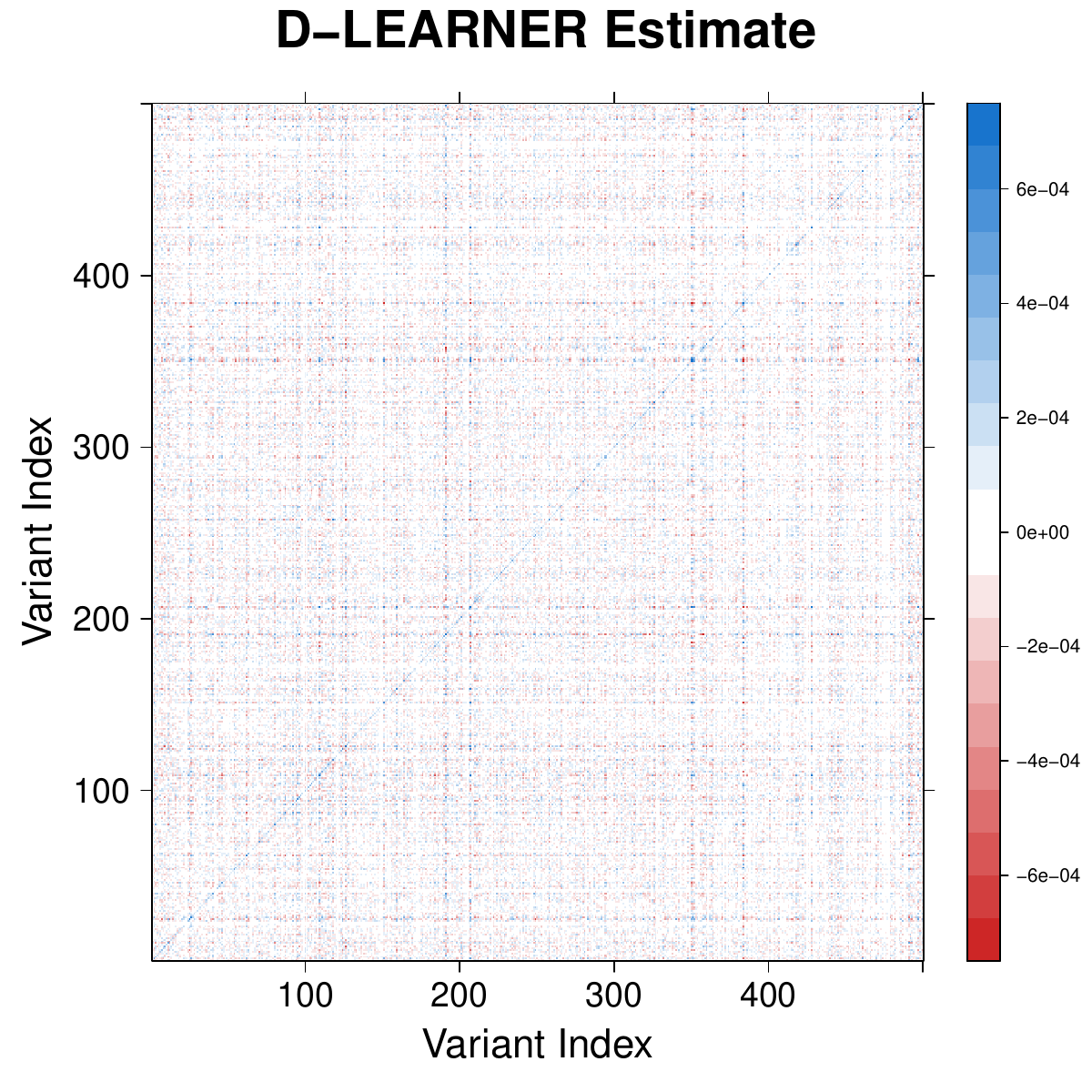}
    \end{subfigure}%
    
    \begin{subfigure}{0.4\textwidth}
        \centering
        \includegraphics[width=\textwidth]{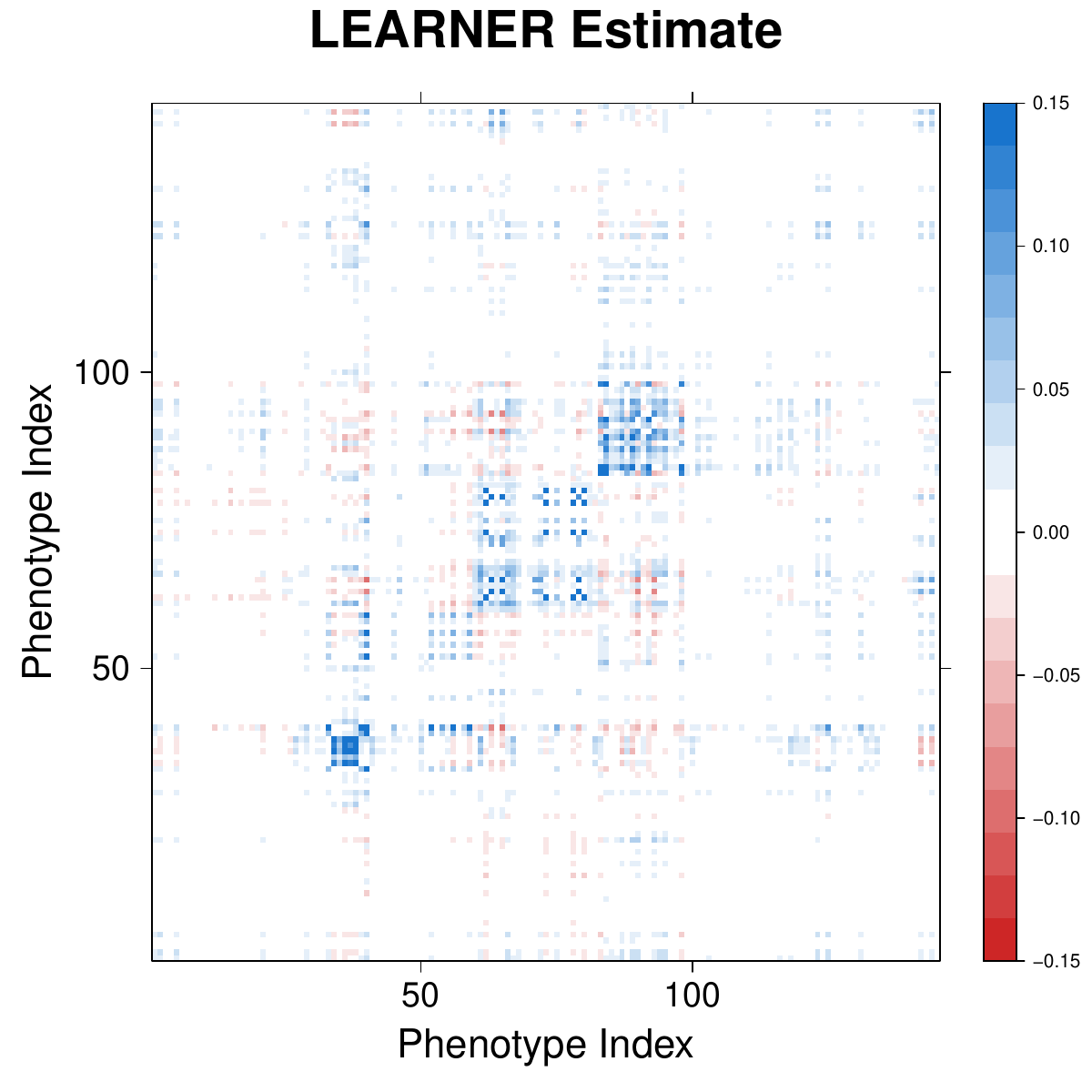}
    \end{subfigure}
    ~ 
    \begin{subfigure}{0.4\textwidth}
        \centering
        \includegraphics[width=\textwidth]{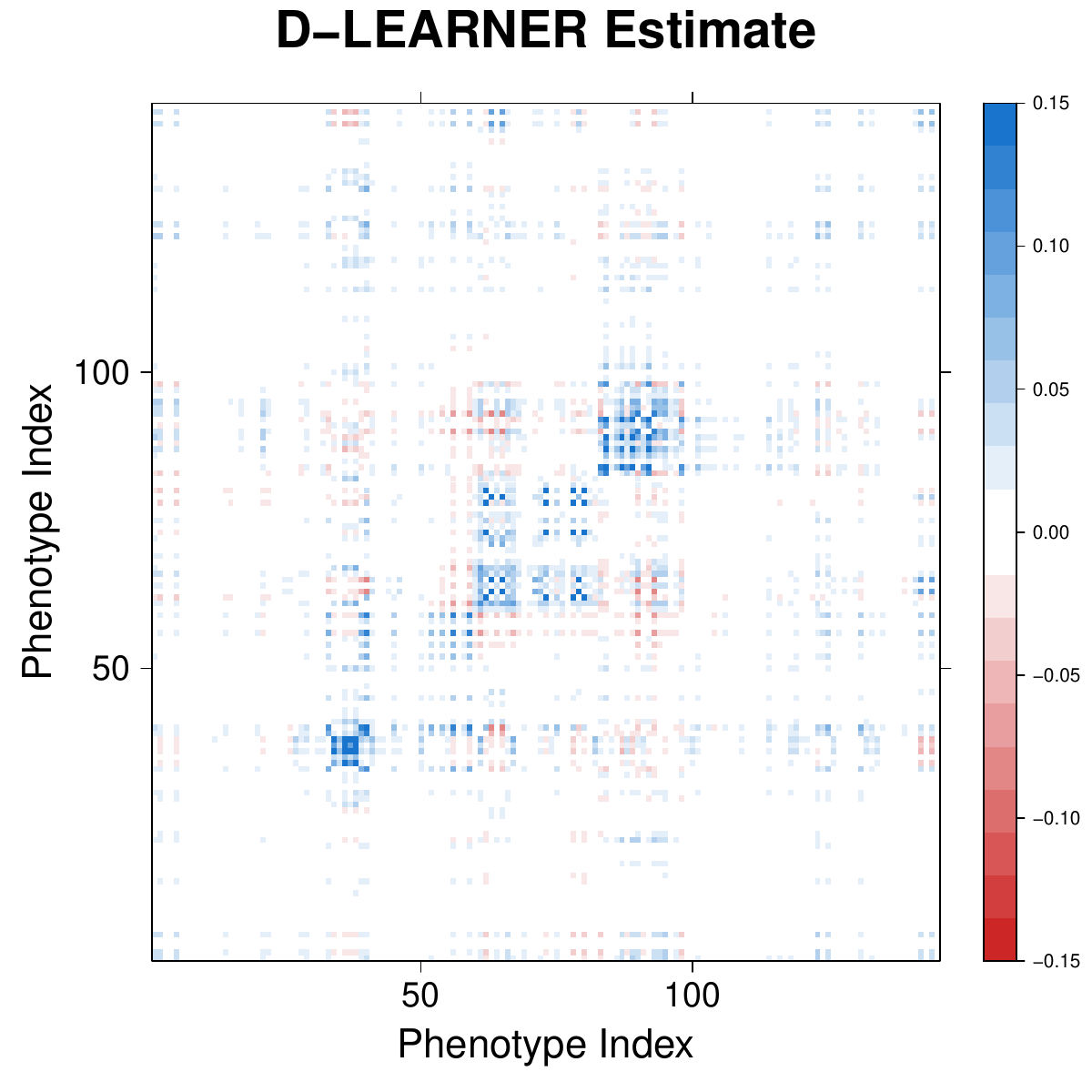}
    \end{subfigure}
    \caption{Heatmaps of 500 randomly selected subset of rows and columns of $\mathcal{P}(\hat{U}_0^{\mathrm{LEARNER}})$ (top left panel) and $\mathcal{P}(\hat{U}_0^{\mathrm{D-LEARNER}})$ (top right panel) as well as the full matrix $\mathcal{P}(\hat{V}_0^{\mathrm{LEARNER}})$ (bottom left panel) and $\mathcal{P}(\hat{V}_0^{\mathrm{D-LEARNER}})$ (bottom right panel). The phenotypes are ordered based on their ICD-10 category, and the 500 selected variants are ordered based on their chromosome and position number. \label{fig: PU PV learner}}
\end{figure}

\begin{figure}[h!]
    \centering
    \begin{subfigure}{0.4\textwidth}
        \centering
        \includegraphics[width=\textwidth]{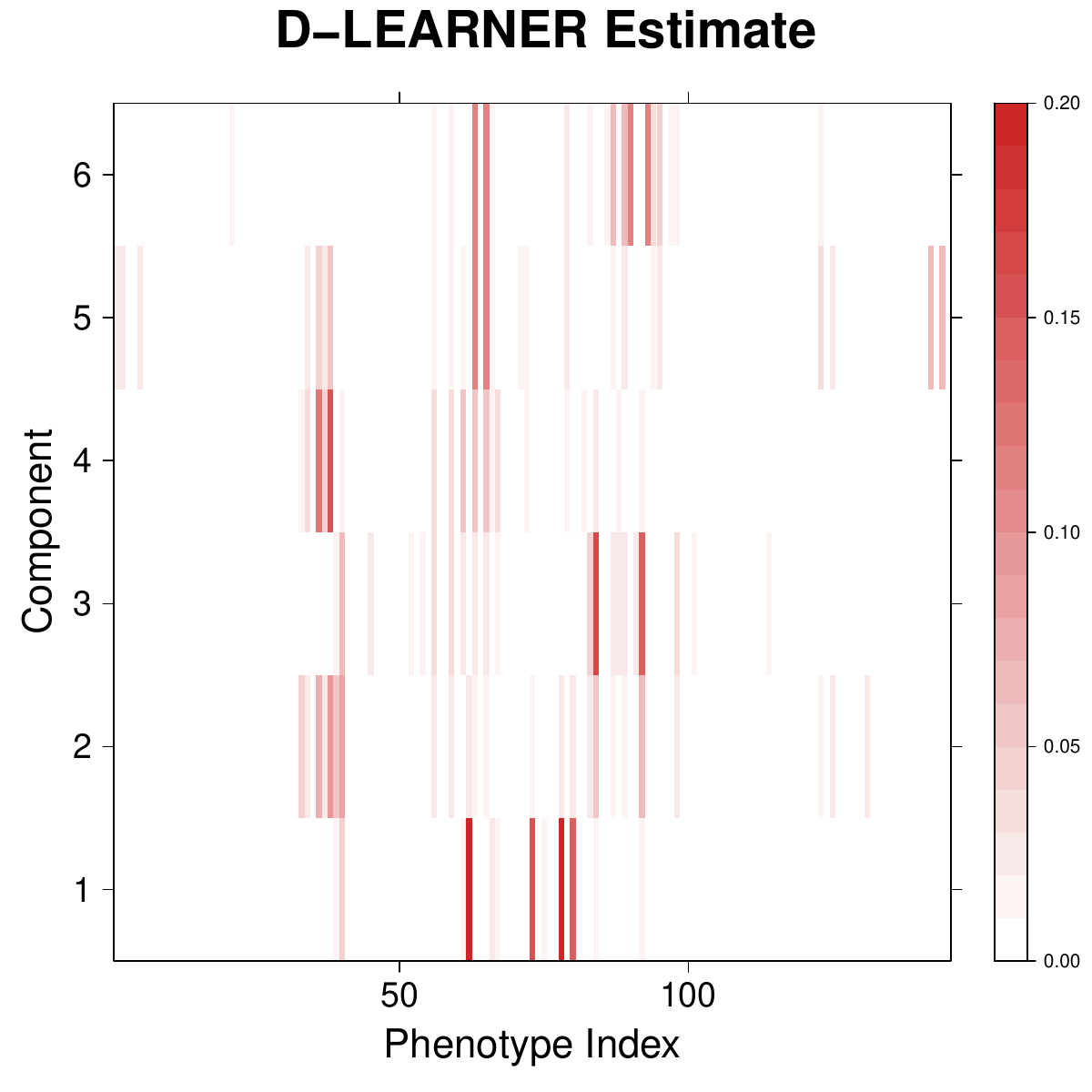}
    \end{subfigure}%
    ~ 
    \begin{subfigure}{0.4\textwidth}
        \centering
        \includegraphics[width=\textwidth]{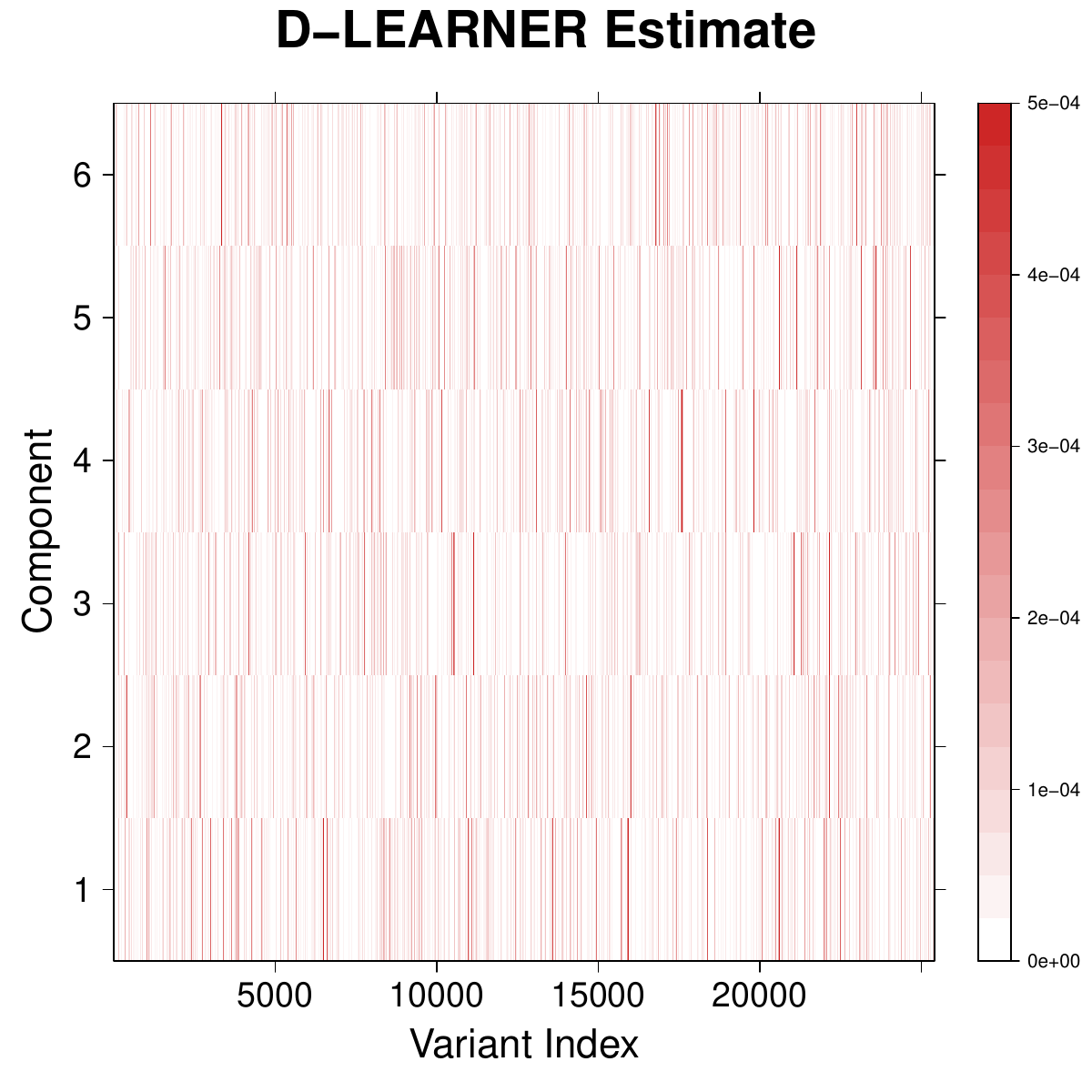}
    \end{subfigure}
    
    \caption{Heatmaps of the matrix of phenotype (left panels) and variant (right panels) contribution scores in the target population based on LEARNER. The phenotypes are ordered based on their ICD-10 category, and the variants are ordered based on their chromosome and position number.\label{fig: contribution learner}}
\end{figure}

\begin{table}[h!]
\caption{Top phenotypes in each latent factor in the target population based on D-LEARNER. For each latent factor, we list the four phenotypes with the highest contribution scores. The phenotype contribution scores are in parentheses. The phenotypes related to the latent factor characterization are in bold text.\label{tab: key phenotypes target d-learner}}
\begin{center}
\begin{tabular}{p{0.06\textwidth}p{0.25\textwidth}p{0.60\textwidth}} \hline 
            Factor & Characterization & Top phenotypes   \\ \hline
            1 & Angina & \textbf{Angina pectoris (0.26)}, \textbf{stable angina pectoris (0.2)}, myocardial infarction (0.16), \textbf{unstable angina pectoris (0.15)} \\
            2 & Thyroid Disorders & \textbf{Hypothyroidism (0.10)}, Type 2 diabetes (0.08), \textbf{Hashimoto's disease (0.07)}, pediatric asthma (0.06) \\
            3 & Asthma & \textbf{Asthma (0.17)}, \textbf{pediatric asthma (0.14)}, type 2 diabetes (0.06), allergic rhinitis (0.04)  \\
            4 & Thyroid Disorders & \textbf{Hypothyroidism (0.16)}, \textbf{Hashimoto's disease (0.13)}, atrial flutter/fibrillation (0.06), cerebral aneurysm (0.06) \\ 
            5 & Aneurysm & \textbf{Unruptured cerebral aneurysm (0.12)}, \textbf{Cerebral aneurysm (0.12)}, head injury (0.06), Achilles tendon rupture (0.06) \\
            6 & Pulmonary Disorders & \textbf{Interstitial lung disease (0.11)}, \textbf{pulmonary fibrosis (0.11)}, unruptured cerebral aneurysm (0.11), cerebral aneurysm (0.11) \\ \hline
\end{tabular}
\end{center}
\end{table} 

\clearpage
\begin{table}[H]
\caption{Top phenotypes in each latent factor in the target population based on LEARNER after performing the varimax rotation. For each latent factor, we list the four phenotypes with the highest contribution scores. The phenotype contribution scores are in parentheses. The phenotypes related to the latent factor characterization are in bold text.}
\begin{center}
\begin{tabular}{p{0.06\textwidth}p{0.25\textwidth}p{0.60\textwidth}} \hline 
            Factor & Characterization & Top phenotypes   \\ \hline
            1 & Angina & \textbf{Angina pectoris (0.27)}, \textbf{stable angina pectoris (0.24)}, myocardial infarction (0.19), \textbf{unstable angina pectoris (0.16)} \\
            2 & Allergic/Respiratory & \textbf{Asthma (0.23)}, \textbf{allergic rhinitis (0.13)}, \textbf{pediatric asthma (0.12)}, \textbf{pollinosis (0.11)} \\
            3 & Diabetes and Eye Disorders & \textbf{Type 2 diabetes (0.32)}, \textbf{type 1 diabetes (0.08)}, \textbf{iritis (0.07)}, \textbf{uveitis (0.06)}  \\
            4 & Thyroid Disorders & \textbf{Hypothyroidism (0.26)}, \textbf{Hashimoto's disease (0.20)}, \textbf{Graves' disease (0.11)}, \textbf{hyperthyroidism (0.11)} \\ 
            5 & Aneurysm & \textbf{Cerebral aneurysm (0.31)}, \textbf{unruptured cerebral aneurysm (0.24)}, subarachnoid hemorrhage (0.11), ischemic stroke (0.04) \\
            6 & Pulmonary Disorders & \textbf{Chronic obstructive pulmonary disease (0.11)}, \textbf{interstitial lung disease (0.09)}, \textbf{chronic bronchitis (0.08)}, \textbf{pneumonia (0.08)} \\ \hline
\end{tabular}
\end{center}
\end{table}

\begin{table}[H]
\caption{Top phenotypes in each latent factor in the target population based on D-LEARNER after performing the varimax rotation. For each latent factor, we list the four phenotypes with the highest contribution scores. The phenotype contribution scores are in parentheses. The phenotypes related to the latent factor characterization are in bold text.}
\begin{center}
\begin{tabular}{p{0.06\textwidth}p{0.25\textwidth}p{0.60\textwidth}} \hline 
            Factor & Characterization & Top phenotypes   \\ \hline
            1 & Angina & \textbf{Angina pectoris (0.26)}, \textbf{stable angina pectoris (0.20)}, myocardial infarction (0.17), \textbf{unstable angina pectoris (0.15)} \\
            2 & Diabetes and Eye Disorders & \textbf{Type 2 diabetes (0.19)}, \textbf{iritis (0.11)}, \textbf{uveitis (0.11)}, \textbf{type 1 diabetes (0.07)} \\
            3 & Allergic/Respiratory & \textbf{Asthma (0.26)}, \textbf{pediatric asthma (0.23)}, \textbf{allergic rhinitis (0.07)}, \textbf{pollinosis (0.06)}  \\
            4 & Thyroid Disorders & \textbf{Hypothyroidism (0.31)}, \textbf{Hashimoto's disease (0.24)}, \textbf{Hyperthyroidism (0.09)}, \textbf{Graves' disease (0.08)} \\ 
            5 & Aneurysm & \textbf{Unruptured cerebral aneurysm (0.31)}, \textbf{cerebral aneurysm (0.31)}, subarachnoid hemorrhage (0.07), Head injury (0.04) \\
            6 & Pulmonary Disorders & \textbf{Pneumonia (0.08)}, \textbf{chronic obstructive pulmonary disease (0.07)}, \textbf{interstitial lung disease (0.07)}, \textbf{pulmonary Fibrosis (0.07)} \\ \hline
\end{tabular}
\end{center}
\end{table}

\bibliographystyle{unsrt}
\bibliography{ref-cur}